\documentclass[lettersize,journal]{IEEEtran}
\usepackage{amsmath,amsfonts}
\usepackage{algpseudocode}
\usepackage{algorithm}
\usepackage{array}
\usepackage[caption=false,font=normalsize,labelfont=sf,textfont=sf]{subfig}
\usepackage{textcomp}
\usepackage{stfloats}
\usepackage{url}
\usepackage{verbatim}
\usepackage{graphicx}
\usepackage{cite}
\usepackage{tcolorbox}
\usepackage{float} 
\usepackage{placeins}
\usepackage{tabularx}
\usepackage{siunitx}
\usepackage{booktabs}
\usepackage{ragged2e}
\usepackage{xspace}
\usepackage{multirow}
\usepackage[numbers]{natbib}
\usepackage[inline]{enumitem}

\setlist[itemize]{label=\textbullet}

\tcbuselibrary{listingsutf8}
\newtcolorbox[auto counter, number within=section]{promptbox}[2][]{%
  colback=white!10!white, colframe=blue!75!black,
  fonttitle=\bfseries, title=Prompt \thetcbcounter: #2,#1
}
\usepackage{tcolorbox}
\tcbuselibrary{skins}
\tcbuselibrary{listings}

\newcommand{\supp}[1]{\textcolor{black}{#1}}
\long\def\rev#1{#1}
\long\def\tnew#1{#1}

\newcommand{\ciciot}{{CICIoT2023}\xspace}
\newcommand{\iot}{{IoT-23}\xspace} 
\newcommand{\siam}{{SiamXBERT}\xspace} %

\newcommand{\subhead}[1]{\vspace {0.04in}\noindent{\textbf{#1.}}}
\usepackage{siunitx}
\tcbuselibrary{skins}
\usepackage{tcolorbox}
\tcbuselibrary{breakable}

\newtcolorbox{summarybox}[2][]{
  colback=gray!8,           
  frame hidden,             
  coltitle=black,
  enhanced,
  breakable,
  sharp corners,
  boxrule=0pt,              
  boxsep=1pt,
  top=2pt, 
  bottom=2pt, 
  left=4pt,                 
  right=4pt,
  attach title to upper,
  after title={:\enskip}, 
  fonttitle=\bfseries,
  title=#2,
  #1
}

\usepackage{orcidlink}
\usepackage{hyperref}
\usepackage[nameinlink,noabbrev]{cleveref}

\begin{document}

\title{\rev{Data-Efficient} Unknown Attack Detection in IoT Networks Using \rev{Pretrained} Language Models
}


\author{Shan Ali$^{\orcidlink{0009-0007-3211-7691}}$,
        Feifei Niu$^{\orcidlink{0000-0002-4123-4554}}$,
        Paria Shirani$^{\orcidlink{0000-0001-5592-1518}}$,
        Lionel C. Briand$^{\orcidlink{0000-0002-1393-1010}}$,~\IEEEmembership{IEEE Fellow}

        
        \thanks{Shan Ali, Feifei Niu and Paria Shirani are with  the School of Electrical Engineering and Computer Science, University of Ottawa, Canada (e-mail: shan.ali@uottawa.ca; feifeiniu96@gmail.com; pshirani@uottawa.ca)}
        \thanks{Lionel C. Briand is with the School of Electrical Engineering and Computer Science, University of Ottawa, Canada, and the Research Ireland Lero Centre, University of Limerick, Ireland (email:lbriand@uottawa.ca)}
}



\maketitle




\begin{abstract}
The rapid evolution of cyberattacks continues to drive the emergence of unknown (zero-day) threats, posing significant challenges for network intrusion detection systems in Internet of Things (IoT) networks. Existing machine learning and deep learning approaches typically rely on large labeled datasets, payload inspection, or closed-set classification, limiting their effectiveness under data scarcity, encrypted traffic, and distribution shifts. Consequently, detecting unknown attacks in realistic IoT deployments remains difficult.
To address these limitations, we propose \siam, \rev{a data-efficient, open-set} Siamese meta-learning framework \rev{based on a pretrained} transformer-based language model for unknown\rev{-}attack detection. The \rev{framework} constructs a dual-modality feature representation by integrating flow-level and packet-level information, enabling richer behavioral modeling while remaining compatible with encrypted traffic. 
\rev{The model learns similarity patterns for known benign and attack classes using \num{100} labeled samples per class. It flags traffic that does not sufficiently match any known class as potentially unknown, using a threshold set without unknown-attack data. 
We evaluate \siam on two IoT intrusion detection datasets against five state-of-the-art detection methods. When training and testing are performed on the same dataset at this labeling budget, \siam attains the highest unknown F1-score, exceeding the strongest competing method on each dataset by \num{84.8}\% and \num{37.4}\%; the improvement is statistically significant against all five methods on one dataset and against four of the five on the other. When a model trained on one dataset is tested on another, \siam transfers effectively in one direction but degrades in the reverse, reflecting the rejection behavior learned from the source traffic rather than the choice of threshold.}
These results highlight the practicality of \siam for \rev{data-efficient} unknown attack detection in real-world IoT environments. 
\end{abstract}


\section{Introduction}
The Internet of Things (IoT) has drawn significant research interest 
due to its widespread applications across several domains such as smart homes~\cite{alaa2017review}, transportation~\cite{zantalis2019review}, 
smart cities~\cite{nassar2018survey}, 
5G environment~\cite{cheng2018industrial}, automotive industry~\cite{krasniqi2016use} and healthcare~\cite{darshan2015comprehensive}. IoT devices are projected to grow from $19.8$ billion in 2025 to $\approx{31.2}$ billion by 2030~\cite{statista}. 
Nevertheless, this growing popularity and rapid development of interconnected IoT devices 
pose significant security risks for both manufacturers and users~\cite{yang2017survey, wang2022survey}. 
The diverse, resource-constrained IoT landscape and inconsistent security standards drive surging threats~\cite{nie2024m2vt}.
For instance, in the first half of 2021, Kaspersky 
recorded more than $1.5$ billion IoT attacks~\cite{kaspersky}, causing significant economic losses and privacy breaches. 

To address this challenge, numerous approaches based on traditional Machine Learning (ML)~\cite{elshrkawey2021enhanced, 
al2023comparative} and Deep Learning (DL)~\cite{gyamfi2022malware,  kilincer2023automated, nie2024m2vt} 
techniques have been proposed for IoT intrusion detection. While these models achieve strong performance in detecting attacks observed during training---referred to as \textit{known attacks}---they often struggle to identify \textit{unknown attacks}~\cite{bou2013cyber} that deviate from previously seen patterns. Such unknown attacks have become increasingly prevalent and consistently rank among the leading causes of security incidents in recent years~\cite{passeri2023q2}. 

However, detecting unknown attacks (including zero-day attacks not observed during training) in IoT environments remains challenging due to evolving threats, scarce labeled data, encrypted or payload-less traffic, and distribution shifts across deployments~\cite{matejeksafe}.
Moreover, IoT environments feature heterogeneous device behaviors, diverse communication protocols, and highly dynamic traffic patterns, which introduce substantial variability and further complicate the detection of unseen attacks~\cite{martins2025adaptive}. As a result, unknown attack detection remains one of the most critical and challenging problems in modern network security.

Despite extensive research, existing solutions share several fundamental limitations that hinder the practical detection of unknown attacks. (i) Most approaches depend heavily on large labeled training sets to learn reliable decision boundaries, often requiring thousands of samples per class or substantial training splits, which is unrealistic for emerging or rare attacks~\cite{zha2025dm,zhao2022can,zhong2024rfg,li2024ids,alsuwaiket2025zeroday}. For example, IDS-Agent relies on the convergence of multiple ML models trained on \num{10}\% of the dataset, corresponding to approximately \num{4.6} million samples~\cite{li2024ids}. 
(ii) Many methods rely on raw payload inspection~\cite{zha2025dm,farrukh2024ais,min2018tr}, which becomes ineffective in encrypted IoT traffic and raises privacy concerns. 
(iii) Detectors are commonly optimized under closed-set or intra-dataset settings and therefore exhibit limited generalization to unseen behaviors or cross-dataset distribution shifts~\cite{wang2025lapis,qiu2025disentangled,zha2025dm}. 
Although recent hybrid~\cite{alsuwaiket2025zeroday} and LLM-assisted~\cite{li2024ids,saheed2025autonomous} systems introduce reasoning or orchestration mechanisms, they still depend on large labeled downstream training data and do not inherently address \rev{data-efficiency or support} open-set generalization. 
Collectively, these limitations highlight a gap for payload-free, data-efficient, and robust methods that can reliably detect unknown attacks under realistic distribution shifts.

To overcome these challenges, we introduce \siam, a \rev{data-efficient, open-set} approach for detecting unknown attacks. Instead of learning closed-set decision boundaries, \siam adopts a meta-learning strategy modeling similarity relationships between benign and attack behaviors. By leveraging a Siamese architecture with a domain-specific SecBERT~\cite{secbert} backbone (a transformer encoder pretrained on large-scale cybersecurity corpora), the model learns transferable, semantically rich traffic embeddings that generalize to unseen attack patterns with a small number of labeled samples. It operates on flow-level and packet-level (i.e., header) features, ensuring compatibility with encrypted traffic, and performs threshold-based open-set inference to explicitly detect unknown attacks. This design improves generalization in \rev{data-efficient} and cross-dataset settings without relying on payload information, underscoring its practicality in real-world environments.

Our key contributions are summarized as follows:
\begin{itemize} [leftmargin=.32cm,noitemsep,topsep=0pt]
    \item We conduct a systematic evaluation of ML-, DL-, and \rev{BERT-based} approaches for known IoT attack detection across both small- and large-scale training settings, 
    \rev{characterizing their} data efficiency, scalability, and robustness. 

    \item We propose \siam, a novel approach that integrates a Siamese network with SecBERT to detect unknown IoT attacks. \rev{\siam learns similarities between benign and known attack traffic samples from dual-modality features. It calibrates a threshold for detecting unknown attacks by temporarily treating selected known attack classes as unknown, without using data from the actual unknown attacks.
    }
   \rev{This integration of Siamese learning, triplet loss, SecBERT-based feature representations, and pseudo-unknown threshold calibration provides a unified framework for data-efficient unknown-attack detection.} 


    \item We systematically benchmark \siam against \rev{five} state-of-the-art (SOTA) \rev{baseline methods} for unknown attack detection \rev{in two evaluation settings: \textit{within-dataset}, where training and testing use the same dataset, and \textit{cross-dataset}, where a model trained on one dataset is tested on another}. 
    \rev{Using \num{100} labeled samples per known class, \siam achieves the highest unknown F1-score on both datasets in a within-dataset setting; over ten runs, the improvement is statistically significant against all five baseline methods on one dataset and against four of the five on the other. In a cross-dataset setting, \siam transfers effectively in one direction but degrades in the reverse, reflecting the rejection behavior learned from the source traffic rather than the choice of threshold.}

    \item To facilitate reproducibility and future research, we release our datasets, implementations of \siam, and baseline models as an open-source replication package 
\end{itemize}

The remainder of this paper is organized as follows. 
Section~\ref{sec:related_work} reviews related work on unknown-attack detection in IoT networks. Section~\ref{sec:proposed_approach} introduces the proposed approach, \siam. Section~\ref{sec:study_design} describes the study design. Section~\ref{sec:results} presents the results. Section~\ref{sec:discussion} discusses the findings and threats to validity. Finally, Section~\ref{sec:conclusion} concludes the paper with future work.

\section{Related Work}\label{sec:related_work}

\tnew{We review prior studies on unknown-attack detection in IoT networks, covering ML-, DL-, and LLM-based methods, including few-shot learning and dual-modality flow- and packet-level feature extraction. We then summarize recurring limitations that motivate our approach.}

\subhead{\tnew{ML- and DL-based approaches}}
\tnew{Traditional ML detects anomalous flows using statistical or distance-based criteria; while lightweight and interpretable, it struggles with unseen attacks~\cite{qiu2025disentangled}. LAPIS~\cite{wang2025lapis} combines supervised classifiers with adaptive clustering to detect unknown attacks by feature-space distance, but flags only the first instance as novel, assigning subsequent samples to a new cluster, limiting sustained detection of recurring zero-day behaviors. Multimodal-SV~\cite{kiflay2024network} uses soft-voting Random Forest (RF) on flows and the first \num{32} payload bytes but detects only known threats. ML also requires substantial training data for reliable decision boundaries and often misses fine-grained encrypted-stream semantics. DL captures richer spatiotemporal dependencies: DM-IDS~\cite{zha2025dm} fuses flow and payload features through an attention-CNN using binary vectors for cybersecurity semantics, requiring \num{50400} training samples; SAFE-NID~\cite{matejeksafe} uses a normalizing-flow safeguard over a lightweight encoder-only transformer for Out-Of-Distribution (OOD) rejection; ACGAN~\cite{zhao2022can} detects unknown classes from in-vehicle CAN ID sequences using an auxiliary classifier GAN; and RFG-HELAD~\cite{zhong2024rfg} isolates unknown attack distributions through contrastive learning and an ensemble of DNNs, GANs, and Deep kNN. Despite their performance, DL methods require thousands of samples per class (\num{10000} for ACGAN) or up to \num{75}\% of the dataset, making them impractical for few-shot emerging IoT threats.}

\subhead{\tnew{\rev{Pretrained encoder- and metric-based methods}}}
\tnew{\rev{Recent methods detect unknown traffic through learned representation distances rather than closed-set classification. Pretrained encoder methods include UT-PAB~\cite{cho2026prototypical}, which aligns class prototypes (single representative embeddings) over BERT representations, assigns samples to the nearest prototype, and rejects those insufficiently close to it. Pretraining with contrastive learning also supports few-shot encrypted-traffic classification and novelty detection~\cite{pretrain2026contrast}. Both rely on raw payload bytes unavailable in encrypted IoT traffic.

Metric-based methods instead learn representations from scratch using contrastive or distance objectives, also found in DL detectors such as RFG-HELAD~\cite{zhong2024rfg}, but use learned distances for rejection. CLOSR~\cite{wilkie2026contrastive} embeds tabular flow features, models known classes as von Mises--Fisher distributions in class-specific projection subspaces, and rejects flows by nearest-class distance. Related methods use online triplet mining with limited labels~\cite{wilkie2026triplet}, few-shot representation comparison for attack classification~\cite{bovenzi2024classifying}, or few-shot incremental learning with IoT open-set rejection~\cite{ma2026openset}. Despite small label budgets, these methods require at least one labeled example of each novel class.
}}

\subhead{\tnew{Hybrid and LLM-based approaches}}
\tnew{Hybrid methods combine DL feature extraction with efficient ML classification. TR-IDS~\cite{min2018tr} uses Text-CNN payload features and flow statistics for RF classification. AIS-NIDS~\cite{farrukh2024ais} converts raw packets into RGB images for 2D-CNN spatiotemporal extraction and open-set recognition of unknown attacks. DIDS-MFL~\cite{qiu2025disentangled} addresses encrypted-traffic data scarcity using multi-scale few-shot learning and graph diffusion but requires balanced sets of \num{200}--\num{1000} samples. XG-NID~\cite{farrukh2025xg} combines dual modalities via a heterogeneous graph neural network and Llama 3 for remedial actions, requiring \num{20000} training samples. LLM-based methods add reasoning: IDS-Agent~\cite{li2024ids} uses GPT-4o to reason over six ML classifiers and take actions, achieving \num{0.61} zero-day recall; Anomaly-Agent~\cite{saheed2025autonomous} extends this to Industrial IoT using open-weight DeepSeek R1 reasoning and SHAP-guided feature selection, achieving \num{0.65} unknown-attack recall. ZeroDay-LLM~\cite{alsuwaiket2025zeroday} uses an edge-cloud architecture and LoRA-adapted, fine-tuned BERT for deep semantic reasoning over sequences of payload bytes and entropy patterns. However, these approaches require large training sets: approximately \num{840000} samples across three intrusion detection datasets and a custom IoT dataset for ZeroDay-LLM~\cite{alsuwaiket2025zeroday}, and \num{4.6} million for IDS-Agent~\cite{li2024ids}. They also face an ``overconfidence trap'': classifiers confidently mislabel unknown traffic as benign, bypassing unknown-attack detection~\cite{li2024ids}.}

\subhead{Limitations and motivation for our work}
Existing work suffers from three critical limitations: (i) high dependence on large-scale training data 
as most methods require hundreds to tens of thousands of samples per class (e.g., $\sim$8K-18K) or even millions of total training samples (e.g., $\sim$0.8M-4.6M) to achieve stable performance~\cite{zha2025dm,wang2025lapis,zhao2022can,zhong2024rfg,li2024ids,alsuwaiket2025zeroday,saheed2025autonomous,matejeksafe}\rev{; the few-shot and metric-learning methods above are the exception, but they still require a labeled example of each novel class, which our setting does not provide}; (ii) reliance on raw payloads that are unavailable in encrypted IoT traffic~\cite{zha2025dm}; and (iii) evaluation restricted to intra-dataset holdouts, failing to account for real-world distribution shifts~\cite{matejeksafe}.
We address these gaps by fine-tuning a domain-specific SecBERT model with only 100 samples per class (i.e., \num{500} and \num{2500} samples, respectively), achieving extreme data efficiency while maintaining payload-free resilience in cross-dataset scenarios. By fusing flow (Zeek~\cite{zeek2024}) 
and packet (DPKT~\cite{dpkt2022})
modalities, our framework remains payload-free and efficient. We also demonstrate our model's superiority by conducting explicit cross-dataset evaluations.


\section{Proposed Approach}
\label{sec:proposed_approach}

\begin{figure*}[tb]
    \centering
    \includegraphics[width=0.80\textwidth]{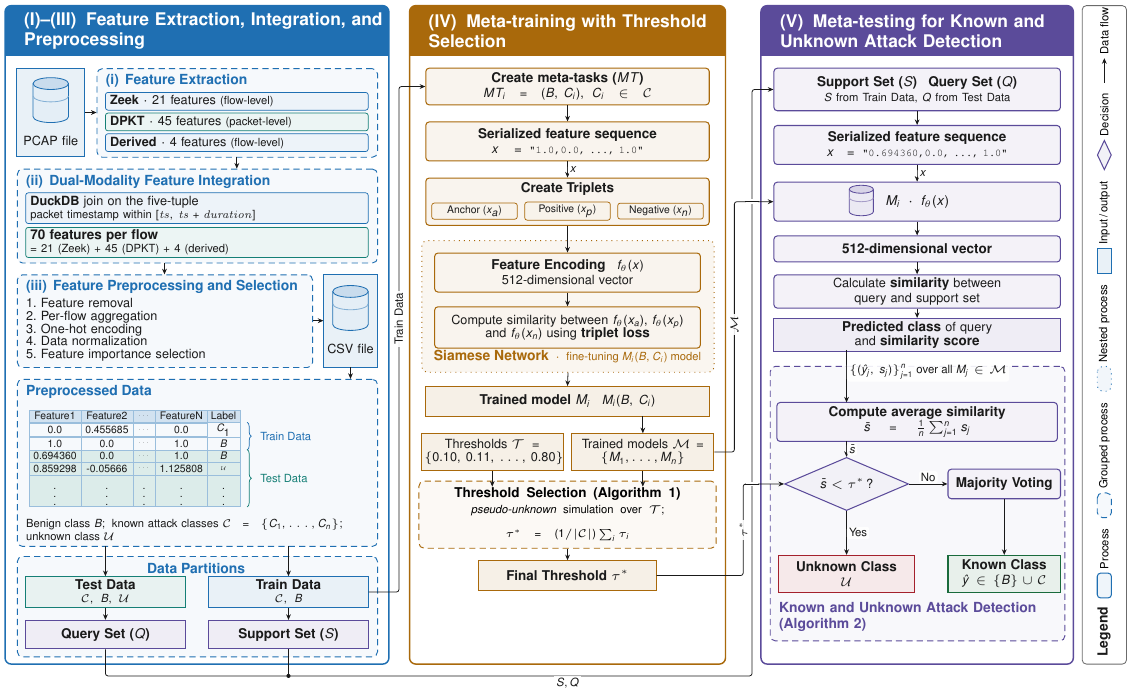}

    \vspace{-0.5cm}
    \caption{\rev{End-to-end architecture of \siam: \textbf{(I)--(III)} feature extraction, integration and preprocessing, \textbf{(IV)} meta-training and threshold selection, and \textbf{(V)} meta-testing. 
    }}
    \label{fig:system_overview}
    \vspace{-15pt}
\end{figure*}

In this study, we introduce \siam, a novel approach for detecting unknown IoT attacks that integrates a Siamese network with SecBERT~\cite{secbert} and enhances it with comprehensive feature engineering. \siam captures both flow-level and packet-level features from PCAP files to provide a holistic representation of network behavior. 
\rev{As illustrated in Figure~\ref{fig:system_overview}, it consists of five stages: (i) feature extraction, (ii) dual-modality feature integration, (iii) feature preprocessing and selection, (iv) meta-training with threshold selection, and (v) meta-testing for known and unknown attack detection.} 



    
    
    
\subsection{Feature Extraction}
\label{sec:feature_extraction}

\tnew{We extract flow- and packet-level features from raw PCAP files to capture both global communication behavior and fine-grained traffic dynamics. Motivated by the discriminative value of directional and rate-normalized features for IoT intrusion detection~\cite{awad2022examining}, we also derive four features capturing temporal intensity and directional asymmetry.}

\subhead{\tnew{Flow-level Feature Extraction}}
\tnew{Using Zeek~\cite{zeek2024}, we extract \num{21} features describing connection timestamps, identifiers, source/destination IP addresses and ports, protocols, duration, byte and packet counts, and service types.}

\subhead{\tnew{Packet-level Feature Extraction}}
\tnew{Using DPKT~\cite{dpkt2022} for efficient packet parsing, we extract \num{39} features following Neto et al.~\cite{neto2023ciciot2023}, supplemented with timestamps, source/destination IP addresses and ports, and flow duration, resulting in a comprehensive set of \num{45} features.}

\subhead{\tnew{Derived Features Extraction}}
\tnew{Following~\cite{awad2022examining}, we derive four features from Zeek flows to capture directional and rate-normalized traffic characteristics:}
\begin{itemize}[leftmargin=.32cm,noitemsep,topsep=0pt]
    \item \tnew{\textit{Byte ratio:}
    $\frac{\text{orig\_bytes}}{\text{resp\_bytes}+1}$
    measures the relative bytes exchanged by the originator and responder, capturing asymmetries associated with malicious activities such as scanning, beaconing, and exfiltration that absolute counts may miss.}

    \item \tnew{\textit{Origin packet rate:}
    $\frac{\text{orig\_pkts}}{\text{duration}+1}$
    measures the originator's packet transmission frequency normalized by \textit{flow duration}, capturing temporal intensity.}

    \item \tnew{\textit{Origin byte rate:}
    $\frac{\text{orig\_ip\_bytes}}{\text{duration}+1}$
    measures the originator's IP-layer byte throughput, complementing packet rate to distinguish high-bandwidth attacks with similar packet counts but substantially larger byte transfers.}

    \item \tnew{\textit{Direction:}
    $\mathrm{orig\_bytes}-\mathrm{resp\_bytes}$
    quantifies the directional byte dominance, distinguishing attacks with similar aggregate volumes but different traffic directions.}
\end{itemize}

\tnew{Flow-level features summarize overall behavior, while packet-level features capture temporal and structural variations indicative of subtle attacks or anomalies. We store features in Comma-Separated Values (CSV) format for scalable querying and aggregation during feature combination using DuckDB\footnote{\footnotesize\url{https://duckdb.org/}}.}

\subsection{Dual-Modality Feature Integration}
\label{sec:feature_combination}

We combine Zeek's flow-level features with DPKT's packet-level features and four derived features to form a unified representation. A flow is characterized by a five-tuple identifier (\texttt{src\_ip}, \texttt{src\_port}, \texttt{dst\_ip}, \texttt{dst\_port}, and \texttt{Protocol\_Type}), representing the source and destination IP addresses and ports, and the protocol.
When matching packets to a flow, we consider any five-tuple matching flow where the packet timestamp falls within the range defined by the flow's start time (\texttt{ts}) and duration (\texttt{duration}): $[ts, ts + \texttt{duration}]$.
This ensures that each packet is correctly associated with the corresponding 
flow~\cite{matejeksafe, matejek2024safeguarding, kiflay2024network}.
This process results in a merged dataset with \num{70} features (\num{21} from Zeek, \num{45} from DPKT, and \num{4} derived features).

\subsection{Feature Preprocessing and Selection}\label{sec:preprocessing}

To improve feature quality, we 
remove redundant features and identifiers from the merged dataset that may lead to overfitting or device-specific learning as follows. 
From the Zeek feature set, we remove timestamp (\texttt{ts}), unique connection identifier (\texttt{uid}), source and destination IP addresses and ports (\texttt{id.orig\_h}, \texttt{id.resp\_h}, \texttt{id.orig\_p}, \texttt{id.resp\_p}), and protocol (\texttt{proto}). Ports and protocol are excluded since equivalent fields already exist in DPKT (i.e., \texttt{src\_port}, \texttt{dst\_port}, \texttt{Protocol\_Type}). From DPKT, we remove \texttt{ts}, \texttt{src\_ip}, and \texttt{dst\_ip} to prevent learning environment-specific IP dependencies rather than traffic behavior.


After removing the aforementioned ten features, the resulting dataset comprises \num{60} distinct, non-redundant features 
reflecting both flow- and packet-level information.  
To aggregate the packet-level features within each flow, we take the mean value for numerical features and the mode (i.e., the most frequently observed value) for categorical features across all packets~\cite{neto2023ciciot2023}.
To ensure applicability in encrypted traffic scenarios, feature extraction relies exclusively on observable outer-layer protocol headers and statistical flow characteristics derived from packet-level parsing (e.g., DPKT) and flow analysis, rather than application-layer payload content. When protocols are encapsulated within encrypted channels, only the externally visible header information and traffic statistics remain available for analysis. Accordingly, features indicating the presence of specific protocols (e.g., HTTP, DNS) are represented as binary (0/1) indicators denoting whether the corresponding protocol is observed within the flow.

Following aggregation, categorical attributes are transformed using one-hot encoding.
Next, numerical features are normalized using MinMax scaling to map all training data values to the range $[0,1]$.
Finally, feature importance-based selection is applied to further reduce feature dimensionality. Specifically, an RF classifier trained on the training set is used to estimate feature importance scores, and a fixed threshold is adopted to discard low-contributing features.



\subsection{Meta-training with Threshold Selection}
\label{sec:meta_threshold}


\rev{\siam trains a collection of class-specific Siamese models to learn similarity relationships between network flows in an embedding space. Each shared-weight branch uses SecBERT as a pretrained transformer encoder and a feed-forward projection to obtain a representation for cosine-similarity computation. The implementation constructs independent binary classification (benign-versus-known-attack) tasks and optimizes each with triplet loss. We use ``meta-task'' and ``meta-training'' here to denote class-wise task decomposition.}

\subsubsection{Meta-training}
\label{sec:meta_training}
The main objective of this module is to learn a transferable similarity-based embedding space 
and yield task-specific models that are subsequently adapted for detecting unknown attacks. 
Given the preprocessed data, we denote the benign class by \(B\), the set of known attack classes by \(\mathcal{C}=\{C_1,\dots,C_n\}\), and the unknown class by 
\(\mathcal{U}\). As shown in the pipeline in Figure~\ref{fig:system_overview}, we first decompose the training process into a collection of binary \emph{meta-tasks} ($MT$), one per attack type:
$MT_i = (B, C_i), C_i \in \mathcal{C}.$
Given \(\mathcal{C}=\{C_1,C_2,C_3\}\) as an example, the meta-training process iterates through tasks \(MT_1, MT_2, MT_3\). For each task $MT_i$, it fine-tunes the Siamese network, resulting in a set of task-specific trained models \(\{M_1, M_2, M_3\}\), 
labeled as ``Trained model $M_i$''. 

\rev{More specifically, for each task \(MT_i\), we select training samples from its two classes ($B, C_i$) and serialize each consistently ordered structured feature vector as a comma-separated sequence. This sequence is the input to the pretrained tokenizer and encoder. The ``\textit{Create Triplets}'' function operates on these serialized feature sequences to create triplets.}
A triplet consists of an \emph{anchor} \(x_a\), a \emph{positive} \(x_p\) from the same class as \(x_a\), and a \emph{negative} \(x_n\) from the other class in the \(MT_i\) task pairing.
These triplets are then processed by the Siamese network. 

The Siamese encoder \(f_\theta(\cdot)\) is implemented using a shared-weight SecBERT model. As depicted in Figure~\ref{fig:system_overview}, SecBERT encodes each \rev{serialized sequence} (\(x_a, x_p, x_n\)) to a fixed-dimensional \rev{representation}, denoted as \(f_\theta(x)\), for similarity learning.
The network is then optimized using a triplet loss function, which computes the similarity between the anchor-positive and anchor-negative embedding pairs, aiming to pull \(f_\theta(x_a)\) and \(f_\theta(x_p)\) closer while pushing \(f_\theta(x_a)\) and \(f_\theta(x_n)\) apart. We use cosine similarity \(s\) as the similarity measure in the embedding space; the loss therefore enforces a high \(s(f_\theta(x_a),f_\theta(x_p))\) and a low \(s(f_\theta(x_a),f_\theta(x_n))\).

\subsubsection{Threshold Selection}
\label{sec:threshold_selection}

\rev{To detect previously unseen attacks, we introduce a decision threshold \(\tau\). It is selected by \emph{pseudo-unknown} calibration, in which each known attack class is alternately withheld and treated as unknown.}
For each known attack class \(C_i\), \(i\in\{1,\dots,n\}\), we withhold \(M_i\) and apply the remaining models to \(C_i\cup B\), treating \(C_i\) as pseudo-unknown, so the calibration set contains the held-out pseudo-unknown class and benign examples.
We test \(\mathcal{T}=\{0.10,0.11,\dots,0.80\}\) at \num{0.01} intervals: known-sample rejection increases from approximately \num{0.1}--\num{0.3}\% at \num{0.10} to \num{90}--\num{94}\% at \num{0.80}, covering boundaries that accept almost all known samples through to those rejecting most.
We average cosine similarities across remaining models and assign samples below \(\tau\) to \(\mathcal{U}\); otherwise, majority voting assigns the most frequent predicted class in \(\{B\}\cup\mathcal{C}\).
We select \(\tau_i\) maximizing the pseudo-unknown class F1-score and compute \(\tau^*=\frac{1}{|\mathcal{C}|}\sum_{i=1}^{|\mathcal{C}|}\tau_i\).
Algorithm~\ref{alg:threshold_selection} summarizes the procedure.

\begin{algorithm}[t]
\small
\caption{Threshold Selection}
\label{alg:threshold_selection}
\begin{algorithmic}[1]
\State \textbf{Input:} Known classes $\mathcal{C} = \{C_1, \dots, C_n\}$, Benign class $B$, Models $\mathcal{M} = \{M_1, \dots, M_n\}$, 
\rev{Thresholds $\mathcal{T} = \{0.10, 0.11, 0.12, \ldots, 0.80\}$}, Unknown label $\mathcal{U}$
\State \textbf{Output:} Final threshold $\tau^*$
\For{each $C_i \in \mathcal{C}$}
    \State Treat $C_i$ as \emph{pseudo-unknown}
    \For{each $\tau \in \mathcal{T}$}
         \State $\widehat{\mathcal{Y}}_{i,\tau} \leftarrow \emptyset$ \Comment{Store predicted labels for this $(i,\tau)$}
        \For{each sample \rev{$x_k \in C_i \cup B$}} 
            \State $(s_{j,k},\hat{y}_{j,k}) \leftarrow M_j(x_k)\;\;\forall j\neq i$ \Comment{$s_{j,k}$: similarity, $\hat{y}_{j,k}$: predicted label}
      

            \State $\bar{s}_k \leftarrow \frac{1}{n-1}\sum_{j\neq i} s_{j,k}$ \Comment{Similarity average}

            \State $\tilde{y}_k \leftarrow \mathrm{mode}(\{\hat{y}_{j,k}:j\neq i\})$ \Comment{Majority vote}
            
            \If{$\bar{s}_k < \tau$} \Comment{Prediction}
            \State $\hat{y}_k \leftarrow \mathcal{U}$ \Else \State $\hat{y}_k \leftarrow \tilde{y}_k$ \EndIf
            \State $\widehat{\mathcal{Y}}_{i,\tau} \leftarrow \widehat{\mathcal{Y}}_{i,\tau} \cup \{\hat{y}_k\}$ \Comment{Accumulate predictions}
            
        \EndFor
        \State $F_i(\tau) \leftarrow \mathrm{F1}_{\mathrm{unk}}(\widehat{\mathcal{Y}}_{i,\tau})$ \Comment{Compute F1}
    \EndFor
    \State $\tau_i \leftarrow \arg\max_{\tau\in\mathcal{T}} F_i(\tau)$
\EndFor
\State $\tau^* \leftarrow \frac{1}{|\mathcal{C}|} \sum_{i=1}^{|\mathcal{C}|} \tau_i$ \Comment{Average threshold}
\end{algorithmic}
\end{algorithm}


\subsection{Meta-testing for Known and Unknown Attack Detection}
\label{sec:unknown_detection}

In this module, we aim to detect previously unseen attacks at inference time by aggregating similarity-based predictions from the trained meta-models and rejecting low-confidence queries as unknown. Overall, \siam employs a Siamese network, SecBERT-based representations, and cosine similarity to distinguish known from previously unseen attacks using a learned decision threshold.

The \emph{support set} \(S\) is selected from known training data and the \emph{query set} \(Q\) from test data; each query \(x_q \in Q\) and each sample in \(S\) is converted into a serialized feature sequence. As shown in Figure~\ref{fig:system_overview}, each trained model \(\{M_i\}_{i=1}^n\) scores \(x_q\) against \(S\), producing a predicted class and a continuous similarity score for each model and query, and \(x_q\) is assigned to \(\mathcal{U}\) when the average similarity \(\bar{s}\) falls below \(\tau^*\) and to the majority-voted class in \(\{B\}\cup\mathcal{C}\) otherwise. The details of the inference phase are summarized in Algorithm~\ref{alg:unknown_detection}.

\begin{algorithm}[t]
\small
\caption{Known and Unknown Attack Detection}
\label{alg:unknown_detection}
\begin{algorithmic}[1]
\State \textbf{Input:} Support set $S$, Query set $Q$, Models $\mathcal{M} = \{M_1, \dots, M_n\}$, Threshold $\tau^*$, Unknown label $\mathcal{U}$ 
\State \textbf{Output:} Predicted label $\hat{y}$ for each $x_q \in Q$
\For{each $x_q \in Q$}
    \State $(s_j,\hat{y}_j) \leftarrow M_j(x_q)\;\;\forall j\in\{1,\dots,n\}$ \Comment{$s_j$: similarity wrt $S$, $\hat{y}_j$: predicted label}
    \State $\bar{s} \leftarrow \frac{1}{n}\sum_{j=1}^{n} s_j$ \Comment{Average similarity}
    \State $\hat{y} \leftarrow \mathrm{majority\_vote}(\{\hat{y}_j\}_{j=1}^n)$ \Comment{Majority vote}
    \If{$\bar{s} < \tau^*$} \Comment{Prediction}
    \State $\hat{y} \leftarrow \mathcal{U}$ \Else \State $\hat{y} \leftarrow \hat{y}$ \EndIf

\EndFor
\end{algorithmic}
\end{algorithm}

\section{Study Design}
\label{sec:study_design}

\subsection{Research Questions}\label{sec:research_questions}
We aim to answer the following Research Questions (RQs):
\begin{enumerate}[label=\textbf{RQ\arabic*.}] 
    \item \textit{How do \rev{BERT-based models} perform compared to traditional ML and DL models for known IoT attack detection?}
    This RQ aims to empirically investigate the performance of SOTA \rev{BERT-based models} (i.e., SecBERT~\cite{secbert}, SecureBERT Plus~\cite{securebertplus}, and BERT~\cite{devlin2018bert}) in detecting known IoT attacks and to benchmark their effectiveness against ML models (i.e., RF and DT) and DL models (i.e., DNN and LSTM).
    
    \item \textit{How does \siam perform under varying configuration settings for the detection of unknown attacks?}
    This RQ aims to evaluate the performance of \siam across different configurations, including combinations of feature sets, feature-importance selection, and maximum token lengths 
    to identify an optimal configuration for \siam.
    
    \item \textit{How does \siam compare with SOTA methods for unknown-attack detection under a within-dataset evaluation setting?}
    This RQ aims to evaluate the effectiveness of \siam in detecting unknown attacks when training and testing are conducted on the same dataset, and to compare its performance with SOTA baselines (i.e., ACGAN~\cite{zhao2022can},  \rev{CLOSR~\cite{wilkie2026contrastive},} SAFE-NID~\cite{matejek2024safeguarding, matejeksafe}, IDS-Agent~\cite{li2024ids}, and RFG-HELAD~\cite{zhong2024rfg}).
    
    \item  \textit{How does \siam compare with SOTA methods for unknown-attack detection under a cross-dataset evaluation setting?}
    This RQ aims to assess the generalization capability of \siam in detecting unknown attacks \rev{drawn from two datasets}, and to benchmark its \rev{effectiveness} against SOTA methods under distribution shifts.
    
\end{enumerate}

\subsection{Baselines}

\subhead{Baselines for known attack detection (RQ1)}
To comprehensively evaluate existing approaches for IoT attack detection, we select representative baselines that have been consistently reported to achieve strong or SOTA performance in prior studies. Specifically, we include two traditional ML models (i.e., RF and DT) and two DL models (i.e., DNN~\cite{abbas2024evaluating} and DL-BiLSTM~\cite{wang2023lightweight}\tnew{, denoted LSTM hereafter}), which have demonstrated competitive detection accuracy across multiple widely-used IoT intrusion detection datasets, such as \ciciot, \iot, and DS2OS~\cite{li2024ids, neto2023ciciot2023, wang2023lightweight}. 
These models are frequently adopted as strong baselines in recent empirical studies and represent both ensemble-based and sequential learning paradigms for attack detection. \rev{To assess pretrained transformer encoders}, we include BERT~\cite{devlin2018bert} as a general-domain baseline, together with SOTA cybersecurity domain-adapted variants, SecBERT~\cite{secbert} and SecureBERT Plus~\cite{securebertplus}. These domain-specific BERT-based models are pretrained on large-scale cybersecurity corpora and outperform generic language models across a range of security-related tasks, making them well-suited baselines for evaluating the added value of domain-specific pretraining in IoT attack detection. \tnew{RF, DT, DNN, and LSTM use scikit-learn 1.6.0 or follow prior work~\cite{wang2023lightweight, abbas2024evaluating}.}

\subhead{Baselines for unknown attack detection (RQ3 \& RQ4)}\label{sec:baselines_uad}
For a fair, comprehensive evaluation, we compare five SOTA baselines with public implementations covering GAN, probabilistic, DL, and LLM-assisted paradigms:
ACGAN~\cite{zhao2022can}, using conditional generation and auxiliary classification to identify unknown attacks as OOD;
SAFE-NID~\cite{matejek2024safeguarding,matejeksafe}, estimating packet-level class-conditional likelihoods with a transformer-enhanced probabilistic model;
IDS-Agent~\cite{li2024ids}, combining \tnew{LLM} reasoning with traditional classifiers for known and unknown attacks;
RFG-HELAD~\cite{zhong2024rfg}, combining contrastive learning, GANs, and deep kNN for open-set detection;
and CLOSR~\cite{wilkie2026contrastive}, using contrastive open-set recognition with von Mises--Fisher class distributions and flow rejection based on distance to the nearest class direction. These methods have been validated on a variety of widely-used intrusion detection benchmarks and represent the current SOTA in unknown attack detection under open-set or OOD settings, making them strong and diverse baselines for comparison. \tnew{We reproduce ACGAN\footnote{\footnotesize\url{https://github.com/evenchen6/CAN_GAN_Anomaly}}, SAFE-NID\footnote{\footnotesize\url{https://github.com/SRI-CSL/trinity-packet}}, IDS-Agent\footnote{\footnotesize\url{https://github.com/PolyLiYJ/IDS_AGENT}}, RFG-HELAD\footnote{\footnotesize\url{https://github.com/RFG-HELAD/RFG-HELAD}}, and CLOSR\footnote{\footnotesize\url{https://github.com/jackwilkie/CLOSR}} using their official replication packages, default configurations, and original testing procedures unless alternative settings are reported. CLOSR reports ranking metrics rather than unknown-class precision, recall, and F1-score, and so provides no decision threshold; as for RFG-HELAD, we calibrate one on a held-out \num{10}\% split of the known training data at a \num{95}\% known-acceptance point, never using unknown test samples.}

\subsection{Dataset}\label{sec:datasets}
In recent years, various IoT security datasets have been proposed to advance research in this area.
Among them, we select the \ciciot~\cite{neto2023ciciot2023} and \iot~\cite{parmisano2020labeled} datasets, which are widely used in prior works~\cite{li2024ids, khorasgani2024empirical, bovenzi2024classifying},
to provide a comprehensive evaluation of the considered techniques across both known and unknown IoT attack scenarios.
Since the original datasets provide only dataset-specific features (\num{39} packet-level features for \ciciot and \num{21} flow-level features for \iot), while the baseline models require different feature granularities, we use Zeek~\cite{zeek2024}
and DPKT~\cite{dpkt2022}
tools to uniformly extract both flow-level and packet-level features from both datasets, enabling all baselines to be evaluated using their original feature representations.

\rev{To prepare the datasets and define the evaluation settings, we proceed as follows. 
(i) Both datasets exhibit severe class imbalance, with class sizes ranging from \num{1252} to \num{71980072} flows in \ciciot and from two to \num{213852924} flows in \iot. Thus, we construct balanced evaluation sets by class-wise stratified sampling. 
(ii) For each dataset we define two \textit{settings}, which we refer to by name throughout the paper. The \textit{known-attack setting} (RQ1) keeps every retained class visible during training. The \textit{unknown-attack setting} (RQ2--RQ4) holds out a disjoint group of attack categories that no method sees during training.
(iii) Since SAFE-NID operates on packets with non-zero payloads, this setting additionally retains a flow only if it carries at least one such packet, filtered with the SAFE-NID replication package~\cite{matejeksafe}, so that packet-based and flow-based methods draw from the same pool. 
\tnew{Following IDS-Agent~\cite{li2024ids}, we then designate the attack classes with the most non-zero-payload flows, along with benign traffic, as known, and use the remaining categories to simulate unknown attacks at test time.}
(iv) Each setting is instantiated at two label budgets. At \textit{small-scale}, every method is trained on \num{100} labeled samples per class, and sampling and training are repeated with the same ten seeds for every method, with results reported as means over the ten runs. At \textit{large-scale}, the known-attack setting uses a fixed, dataset-specific number of samples per class, given below, and the unknown-attack setting uses \num{10}\% of the available flows per known category; large-scale results come from a single run per method and dataset.}

\subhead{Dataset I: \ciciot}
 The \ciciot dataset~\cite{neto2023ciciot2023} is a 
 comprehensive, realistic IoT attack dataset, encompassing \num{33} attacks executed in an IoT topology comprising \num{105} devices. These devices include smart home devices, cameras, sensors, and microcontrollers, effectively mimicking a real-world IoT deployment. The dataset
 contains a total of \num{449144189} samples, with \num{39} packet-level features extracted for each sample. 
 Each sample is labeled as one of the \num{33} attack types or benign traffic. \rev{In the known-attack setting, class-wise sampling retains \num{23} attack classes plus benign traffic, drawing \num{100} samples per class at small-scale and \num{100000} at large-scale. In the unknown-attack setting, the \num{24} 
 attack classes \tnew{with the most non-zero-payload flows}
 together with benign traffic form the \num{25} known categories, and the remaining nine attack categories are held out as unknown. The test set is balanced by construction: \num{18} flows from each known category and \num{50} flows from each unknown category, yielding \num{450} known and \num{450} unknown samples.}

 \subhead{Dataset II: \iot}
 The \iot dataset~\cite{parmisano2020labeled} was collected over a year, providing a comprehensive resource for analyzing IoT network traffic. The dataset comprises \num{20} malware captures and three benign captures, drawn from \num{325244220} records, of which \num{294449255} are labeled as malicious. Each record comprises \num{21} flow-level features and is labeled as one of the \num{15} attack types or as benign traffic, enabling fine-grained analysis of IoT attack detection. \rev{In the known-attack setting, class-wise sampling retains five attack classes plus benign traffic, drawing \num{100} samples per class at small-scale and \num{20000} at large-scale. In the unknown-attack setting, the non-zero-payload filter excludes the \emph{C\&C-PartOfAHorizontalPortScan} and \emph{PartOfAHorizontalPortScan-Attack} classes, as all their packets carry empty payloads; of the remaining classes, the four 
 attack classes \tnew{with the most non-zero-payload flows} together with benign traffic form the five known categories, and the remaining nine attack categories are held out as unknown. The test set comprises \num{189} known samples, drawn as \num{37} flows from each of the four known attack classes plus \num{41} benign flows, and \num{189} unknown samples, taking up to \num{50} flows per unknown category, and all available flows when fewer exist (e.g., \emph{C\&C-Mirai}).}

\subsection{\tnew{Implementation Details}}\label{sec:implementation_details}

\tnew{\subhead{Preprocessing and feature selection}
Beyond the feature removal, one-hot encoding, MinMax normalization, and RF-based selection
described in Section~\ref{sec:preprocessing}, two dataset-specific steps apply. For \ciciot,
which contains only numerical features, we replace missing and undefined values and set
\texttt{Rate} (packets divided by flow duration) to zero when the duration is \num{0.0},
retaining packet count and duration separately. For \iot, following prior
work~\cite{khorasgani2024empirical}, we additionally remove the constant fields
(\texttt{local\_orig}, \texttt{local\_resp}) and map \texttt{tunnel\_parents} values `-' and
`(empty)' to \num{0} and \num{1}, respectively.
RF importance scores are long-tailed on both datasets: \iot has a mean of \num{0.0092}
(75th and 90th percentiles: \num{0.0171} and \num{0.0275}), versus \num{0.00146}
for \ciciot. We therefore use fixed thresholds of \num{0.01} (retaining \num{39} and
\num{35} features for \iot and \ciciot, respectively) and \num{0.02}
(retaining \num{22} and \num{23}).}

\tnew{\subhead{\siam and BERT-based models}
Samples are encoded as comma-separated feature values. For RQ1, the feed-forward
classifier uses a \num{768}-dimensional input, a \num{512}-unit hidden layer
(ReLU, dropout \num{0.1}), and a dataset-specific log-softmax output layer,
trained with negative log-likelihood loss. For RQ2--RQ4, the same projection
maps \num{768}-dimensional embeddings to \num{512} dimensions for \siam,
trained with triplet loss. We use AdamW (learning rate \num{1e-5}),
batch size \num{16}, and early stopping (patience \num{5}), retaining
implementation defaults for unspecified hyperparameters. At inference, cosine
similarity compares each test sample with a training-derived support set of
\num{90} samples per known class. The large-scale \ciciot\ \siam run uses
\num{32} triplets per optimizer step, bfloat16 mixed precision for training
and validation, a maximum token length of \num{384} (longest sequence:
\num{182} tokens; no truncation), and up to \num{15} epochs with the same
early stopping.
All experiments ran on Digital Research Alliance of Canada clusters\footnote{\footnotesize
Narval (NVIDIA A100, \num{8}--\num{40}\,GB) and Fir (NVIDIA H100 SXM5, \num{80}\,GB).}.}

\subsection{Evaluation Metrics}\label{sec:eval_metrics}
We employ standard evaluation metrics for multiclass classification tasks, including Accuracy, Precision, Recall, F1-score, and Area Under the Curve (AUC), which is the area under the Receiver Operating Characteristic (ROC) curve~\cite{lessmann2008benchmarking}. 
For known IoT attack detection (RQ1), 
we report the average Precision, Recall, and F1-score, Accuracy and AUC. For unknown attack detection (RQ2-RQ4), 
we report the weighted average F1-score (W-F1) \rev{over the complete test set, which contains benign, known-attack, and unknown-attack samples.} In addition, for each unknown attack category, \rev{we treat ``unknown'' as the positive class and} report Unknown Precision (U-Pr), Unknown Recall (U-Rc), and Unknown F1-score (U-F1), as well as their average values across all unknown categories. 

\section{Experimental Results}
\label{sec:results}

\subsection{RQ1. 
\rev{BERT-based} Models vs. ML/DL for Known IoT Attack Detection
}\label{sec:results_rq1}
\rev{We evaluate the ML-, DL-, and \rev{BERT}-based models under the known-attack setting of both datasets, at the small-scale and large-scale budgets defined in Section~\ref{sec:datasets}. Each constructed dataset is partitioned into training, validation, and test sets using an $8{:}1{:}1$ split, and every model is trained and evaluated on identical splits.}
Table~\ref{tab:rq1_results} reports the performance of ML-, DL-, and \rev{BERT-based} models on known IoT attack detection under \textit{small-scale} and \textit{large-scale} settings on \ciciot and \iot datasets.

\subhead{Small-scale} Under this setting, the relative strengths of different model families vary across datasets.
On the \ciciot dataset, traditional ML-based models, particularly RF, achieve the strongest performance, reaching an F1-score of \num{0.729}. This result exceeds the best-performing \rev{BERT-based model} (SecBERT, F1=\num{0.658}) by \num{10.8}\%. While \rev{BERT}-based models outperform DL models (e.g., SecBERT vs. DNN: +\num{4.8}\% F1), they do not consistently surpass tree-based ensembles when training data is extremely limited, and the classification space is large.

In contrast, on the \iot dataset, \rev{BERT}-based models exhibit more favorable few-shot behavior. SecureBERT+ achieves the highest F1-score (\num{0.590}), outperforming the best ML baseline (RF, F1=\num{0.554}) by \num{6.5}\% and the best DL model (DNN, F1=\num{0.525}) by \num{12.4}\%. BERT further attains the highest accuracy (\num{0.608}) and AUC (\num{0.879}). These results suggest that, in settings with fewer classes, pretrained language representations can better compensate for limited labeled data.
Across both datasets, DL-based models (DNN and LSTM) consistently perform worst under data scarcity, indicating their greater dependence on larger training sets.

\subhead {Large-scale} Under this setting, 
\rev{BERT}-based models consistently achieve the best overall performance across both datasets.
On \ciciot, BERT attains the highest F1-score (\num{0.822}) and AUC (\num{0.993}), closely followed by SecureBERT+. On \iot, SecureBERT+ and SecBERT achieve the top performance (F1=\num{0.651}, AUC $\approx$ \num{0.906}), outperforming all ML- and DL-based baselines.
As shown, traditional ML models (e.g., RF) do not match the performance of \rev{BERT}-based approaches. The relatively small performance gap among BERT, SecureBERT+, and SecBERT further suggests that both large-scale pretraining and domain adaptation contribute to effective representation learning in this regime.

\begin{summarybox}{Answer to RQ1}
\rev{BERT-based} models achieve competitive performance in small-scale settings and consistently outperform ML and DL baselines in large-scale IoT attack detection. While traditional ML models remain effective with limited data, \rev{BERT}-based approaches demonstrate greater scalability and 
robustness for detecting known IoT attacks.
\end{summarybox}

\vspace{-15pt}
\begin{table}[!htbp]
\centering
\caption{
Comparison of ML, DL, and \rev{BERT-based} models for known IoT attack detection under small- and large-scale settings (RQ1).}
\label{tab:rq1_results}
\vspace{-0.1in}
\setlength{\tabcolsep}{2pt}

\resizebox{\columnwidth}{!}{%
\begin{tabular}{@{}l l ccccc | ccccc@{}}
\toprule
& & \multicolumn{5}{c}{\textbf{Small}} & \multicolumn{5}{c}{\textbf{Large}} \\
\cmidrule(lr){3-7} \cmidrule(l){8-12}
\textbf{Data} & \textbf{Model} & \textbf{P} & \textbf{R} & \textbf{F1} & \textbf{AUC} & \textbf{Acc} & \textbf{P} & \textbf{R} & \textbf{F1} & \textbf{AUC} & \textbf{Acc} \\
\midrule

\multirow{7}{*}{\rotatebox{90}{\textbf{\ciciot}}}

& RF & \textbf{0.743} & \textbf{0.731} & \textbf{0.729} & \textbf{0.975} & \textbf{0.731} 
& 0.806 & 0.803 & 0.801 & 0.990 & 0.803 \\

& DT & 0.689 & 0.684 & 0.681 & 0.839 & 0.684 
& 0.774 & 0.773 & 0.772 & 0.931 & 0.773 \\

& DNN & 0.650 & 0.660 & 0.628 & 0.973 & 0.660 
& 0.752 & 0.730 & 0.711 & 0.988 & 0.730 \\

& LSTM & 0.656 & 0.659 & 0.624 & 0.973 & 0.659 
& 0.777 & 0.768 & 0.764 & 0.989 & 0.768 \\

& SecBERT & 0.684 & 0.669 & 0.658 & 0.974 & 0.669 
& 0.818 & 0.817 & 0.816 & \textbf{0.993} & 0.817 \\

& SecureBERT+ & 0.654 & 0.660 & 0.634 & 0.974 & 0.660 
& 0.823 & 0.821 & \textbf{0.822} & \textbf{0.993} & 0.821 \\

& BERT & 0.643 & 0.660 & 0.627 & 0.972 & 0.660 
& \textbf{0.824} & \textbf{0.822} & \textbf{0.822} & \textbf{0.993} & \textbf{0.822} \\
\midrule

\multirow{7}{*}{\rotatebox{90}{\textbf{\iot}}}

& RF & 0.585 & 0.572 & 0.554 & 0.851 & 0.572 
& 0.665 & 0.630 & 0.621 & 0.896 & 0.630 \\

& DT & 0.535 & 0.542 & 0.515 & 0.812 & 0.542 
& 0.650 & 0.614 & 0.603 & 0.863 & 0.614 \\
& DNN & 0.567 & 0.553 & 0.525 & 0.849 & 0.553 
& 0.633 & 0.617 & 0.607 & 0.896 & 0.617 \\

&  LSTM & 0.544 & 0.547 & 0.511 & 0.837 & 0.547 
& 0.609 & 0.617 & 0.600 & 0.892 & 0.617 \\
& SecBERT & 0.586 & 0.587 & 0.567 & 0.874 & 0.587 
& \textbf{0.694} & \textbf{0.659} & \textbf{0.651} & 0.905 & \textbf{0.659} \\
& SecureBERT+ & \textbf{0.631} & 0.605 & \textbf{0.590} & 0.875 & 0.605 
& \textbf{0.694} & \textbf{0.659} & \textbf{0.651} & \textbf{0.906} & \textbf{0.659} \\

& BERT & 0.618 & \textbf{0.608} & 0.585 & \textbf{0.879} & \textbf{0.608} 
& 0.674 & 0.656 & 0.639 & \textbf{0.906} & 0.656 \\

\bottomrule
\end{tabular}%
}
\vspace{0.1em} 
\parbox{\linewidth}{\scriptsize
\textbf{Abbreviations:} P=Precision; R=Recall; F1=F1-score; Acc=Accuracy}
\vspace{-20pt}

\end{table}

\subsection {
RQ2. 
\siam Effectiveness for Unknown IoT Attack Detection Under Varying Configurations
}
\rev{We evaluate how configuration choices impact \siam's unknown-attack detection performance and identify an effective configuration for each dataset. RQ2 uses the small-scale unknown-attack setting (Section~\ref{sec:datasets}).} 
Under this fixed setup, we systematically vary three key configuration factors of SiamXBERT.
First, for \textit{feature set configurations}, we evaluate 
(i) flow-level features only and packet-level features only, (ii) each of these feature types augmented with our four derived flow-level features, and (iii) the combination of flow-level, packet-level, and our derived features. This allows us to assess the individual and joint contributions of different feature sources.
Second, for \textit{feature importance selection}, we compare three settings: applying feature importance filtering with thresholds of 0.01 and 0.02, and a baseline with no feature importance selection. 
Finally, for \textit{maximum token length}, we vary SecBERT’s maximum input token length (256 vs. 512)
to assess its impact on the performance of unknown-attack detection.

\rev{\subhead{Configuration comparison}}
The performance of \siam under different configuration settings for unknown IoT attack detection is summarized in Table~\ref{tab:rq2_avg_iot23_ciciot2023}.
Overall, no single configuration consistently outperforms the others across both datasets. On the \ciciot dataset, the best performance is achieved by combining flow-level, packet-level\rev{, and derived} features with feature importance selection (threshold 0.01) at a token length of 512. This configuration achieves the highest average unknown F1-score (U-F1) of 0.388 while maintaining a weighted F1-score (W-F1) of 0.641 across all categories, close to the highest observed W-F1 of 0.661.
On the \iot dataset, the best result is obtained using packet-level features with feature importance selection (threshold 0.01) at a token length of 512. This configuration yields the highest U-F1 (0.536) and W-F1 (0.930) across all categories, indicating strong overall detection performance in this setting.
Overall, the results show that the choice of feature sets has a clear impact on the performance of unknown attack detection. 

\rev{\textit{Effect of flow and packet features:}} Single-modality features (\textit{Packet} or \textit{Flow}) yield moderate U-F1 scores on both datasets. Packet-level features tend to achieve higher unknown recall, while flow-level features provide more balanced precision-recall trade-offs, particularly on \ciciot; however, neither modality alone consistently achieves the best performance. Combining flow-level and packet-level features yields consistent improvements in U-F1, particularly on \ciciot, where the \textit{Flow}+\textit{Packet} configuration improves U-F1 by approximately 4-6\% over the best single-modality setting. 

\rev{\textit{Effect of derived features:}} Augmenting feature sets with derived features generally increases unknown recall (often exceeding 10\% points), yet fails to yield consistent U-F1 improvements. 
The recall gains are frequently accompanied by reduced precision, suggesting that derived features may introduce additional noise when used without further filtering. 

\rev{\textit{Effect of feature selection:}} Feature selection has a pronounced positive effect across nearly all configurations, resulting in clear improvements in U-F1 on both datasets. In particular, 
it improves U-F1 by over 30\% on \iot (e.g., \textit{Packet} vs. \textit{Packet}+\textit{FS}) and by roughly 10\% on \ciciot, confirming that removing less informative features significantly improves the robustness of learned representations. 

\rev{\textit{Effect of token length:}} Overall, increasing the token length from 256 to 512 yields no uniform performance gains. 
While some configurations benefit slightly (typically within 2-4\% relative improvement), the overall impact remains limited and configuration-dependent. 
Based on our observations, the best results are achieved with a token length of 512 across datasets. These findings suggest that feature composition and selection have a larger impact on performance than token length.

\rev{\subhead{Ablation study}
We further conduct a controlled ablation study to examine the contributions of the encoder and input representation while holding all other experimental settings fixed. \textit{SecBERT} serves as the reference configuration, where each feature vector is rendered as a comma-separated text sequence before encoding (described in Section~\ref{sec:meta_training}). \textit{BERT} replaces the cybersecurity-domain encoder with a general-domain encoder, thereby isolating the effect of domain-specific pretraining. \textit{Randomly Initialized (Rand Init.)} uses the SecBERT architecture without pretrained weights, thereby isolating the effect of pretraining. All these configurations \textit{serialize} the
feature vector for a pretrained encoder. Finally, \textit{The Tabular Transformer (Transformer)} directly consumes the numerical feature vector rather than a serialized textual representation, thereby isolating the contribution of feature serialization.}

\rev{All four \textit{experimental configurations} in the ablation study share the same setup: the same flow-level, packet-header, statistical, and derived feature set on both datasets, a feature-selection threshold of 0.01, a maximum token length of 512, and the label budget, data splits, held-out categories, and ten random seeds, yielding \(2\) datasets \(\times\) \(4\) experimental configurations \(\times\) \(10\) seeds. The shared feature set and token length are determined by the token budget rather than by measured performance: with a threshold of 0.01, the longest serialized record contains 406 tokens in \ciciot\ and 381 tokens in \iot, ensuring that no sample is truncated. In contrast, without feature selection, every sample exceeds the maximum token length and is truncated. Using the dataset-specific configuration selected in RQ2 would instead evaluate the alternative configurations under settings optimized for SecBERT, potentially biasing the comparison. 
We compare configurations pairwise using seed-level mean scores and the test of Section~\ref{significance_test}.}

\begin{figure}[t]
\centering
\includegraphics[width=\columnwidth]{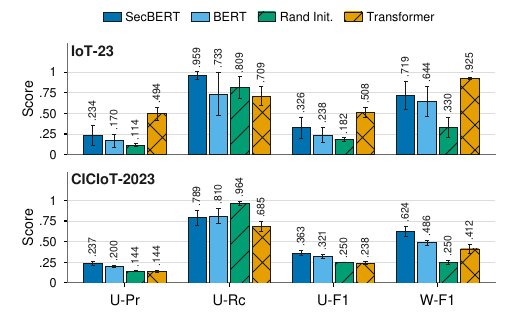}
\vspace{-0.45in}
\caption{\rev{Encoder and input-representation ablation (RQ2). Bars are means over ten seeds, error bars are $\pm$1 standard deviation, and the value above each bar is the mean.}}
\label{fig:rq2_encoder_ablation}
\vspace{-15pt}
\end{figure}

Figure~\ref{fig:rq2_encoder_ablation} shows an ablation study on the encoder and input representation by holding the feature set, label budget, training objective, feature-selection threshold, and random seeds fixed. On \ciciot, SecBERT achieves the highest U-F1 and W-F1. It outperforms general-domain BERT by \(\Delta\mathrm{U\text{-}F1}=+0.042\) (Holm-adjusted \(p = 0.010\)) and Rand Init. by \(+0.113\) (\(p = 0.006\)). Thus, pretraining improves unknown-attack detection, and cybersecurity-domain pretraining provides an additional benefit beyond general-domain pretraining. On \iot, SecBERT likewise achieves higher U-F1 than BERT and Rand Init.; however, only the improvement over Rand Init. is statistically significant (\(+0.144\), \(p = 0.018\)), while its advantage over BERT is inconclusive across ten seeds (\(+0.088\), \(p = 0.252\)). 
The wider error bars indicate higher seed-level variation for the serialized configurations on \iot\ than on \ciciot.

\rev{The comparison with direct numeric encoding differs between datasets. Transformer is the weakest on \ciciot, where SecBERT improves U-F1 by \(0.125\) (\(p = 0.006\)). In contrast, it is the strongest configuration on \iot; SecBERT has a lower U-F1 by \(0.182\) (\(p = 0.029\)) and the Transformer has the smallest W-F1 variation among the four configurations. This difference is reflected in SecBERT's precision--recall trade-off: on \iot, it achieves U-Rc \(=0.959\) but U-Pr \(=0.234\), 
indicating many false positives, whereas on \ciciot\ it achieves a more balanced U-Rc \(=0.789\) and U-Pr \(=0.237\). Overall, the ablation shows that pretraining improves performance relative to random initialization on both datasets, and cybersecurity-domain pretraining further benefits \ciciot. Feature serialization with a pretrained encoder is preferable to native numeric encoding only on \ciciot, which has the larger known-class space and more pseudo-unknown classes for threshold selection.}

\begin{summarybox}{Answer to RQ2}
No configuration consistently outperforms others across both datasets; however, combining multimodal features and feature-importance selection generally improves performance, whereas token length plays a comparatively minor role. \rev{Pretraining contributes on both datasets and domain-specific pretraining contributes on \ciciot, but feature serialization into a pretrained encoder is only preferable to native numeric encoding on the dataset with the larger known-class space.}
\end{summarybox}

\vspace{-0.22in}
\begin{table}[!htbp]
\centering
\caption{Average performance of \siam for unknown attack detection under different configurations (RQ2).}
\label{tab:rq2_avg_iot23_ciciot2023}
\vspace{-5pt}
\setlength{\tabcolsep}{4pt}
\resizebox{\linewidth}{!}{%
\begin{tabular}{l c
                S[table-format=1.3] S[table-format=1.3] S[table-format=1.3] S[table-format=1.3]
                S[table-format=1.3] S[table-format=1.3] S[table-format=1.3] S[table-format=1.3]}
\toprule
\multirow{2}{*}{\textbf{Feature Set} (Threshold)} & \multirow{2}{*}{\textbf{Token}} &
\multicolumn{4}{c}{\textbf{IoT-23}} &
\multicolumn{4}{c}{\textbf{CICIoT-2023}} \\
\cmidrule(lr){3-6}\cmidrule(lr){7-10}
& & \textbf{W-F1} & \textbf{U-Pr} & \textbf{U-Rc} & \textbf{U-F1}
  & \textbf{W-F1} & \textbf{U-Pr} & \textbf{U-Rc} & \textbf{U-F1} \\
\midrule

\textit{Packet} & 256 & 0.866 & 0.336 & 0.812 & 0.406 & 0.455 & 0.185 & 0.834 & 0.302 \\
           & 512 & 0.851 & 0.336 & 0.846 & 0.404 & 0.436 & 0.169 & 0.804 & 0.279 \\

\textit{Packet+FS (0.01)} & 256 & 0.886 & 0.424 & 0.880 & 0.499 & 0.489 & 0.185 & 0.799 & 0.299 \\
                          & \textbf{512} & \textbf{0.930} & \textbf{0.484} & 0.803 & \textbf{0.536} & 0.472 & 0.182 & 0.806 & 0.296 \\

\textit{Packet+FS (0.02)} & 256 & 0.878 & 0.415 & 0.863 & 0.476 & 0.493 & 0.190 & 0.844 & 0.310 \\
                          & 512 & 0.891 & 0.432 & 0.914 & 0.511 & 0.492 & 0.185 & 0.817 & 0.301 \\

\textit{Packet+DE} & 256 & 0.770 & 0.235 & 0.965 & 0.332 & 0.488 & 0.189 & 0.823 & 0.307 \\
              & 512 & 0.806 & 0.239 & 0.864 & 0.326 & 0.489 & 0.189 & 0.798 & 0.305 \\

\textit{Packet+DE+FS (0.01)} & 256 & 0.867 & 0.347 & 0.782 & 0.412 & 0.576 & 0.216 & 0.793 & 0.340 \\
                             & 512 & 0.846 & 0.358 & 0.832 & 0.421 & 0.572 & 0.225 & 0.834 & 0.354 \\
\textit{Packet+DE+FS (0.02)} & 256 & 0.869 & 0.376 & 0.739 & 0.420 & 0.557 & 0.210 & 0.821 & 0.334 \\
                             & 512 & 0.872 & 0.385 & 0.767 & 0.426 & 0.577 & 0.218 & 0.765 & 0.338 \\ \hline

\textit{Flow} & 256 & 0.883 & 0.313 & 0.743 & 0.381 & 0.598 & 0.223 & 0.788 & 0.346 \\
           & 512 & 0.879 & 0.327 & 0.790 & 0.401 & 0.566 & 0.211 & 0.832 & 0.335 \\


\textit{Flow+FS (0.01)} & 256 & 0.904 & 0.367 & 0.855 & 0.448 & 0.610 & 0.216 & 0.776 & 0.336\\
                        & 512 & 0.909 & 0.385 & 0.883 & 0.471 & 0.625 & 0.229 & 0.789 & 0.353 \\
\textit{Flow+FS (0.02)} & 256 & 0.909 & 0.383 & 0.867 & 0.464 & 0.652 & 0.228 & 0.706 & 0.343 \\
                        & \textbf{512} & 0.916 & 0.404 & 0.862 & 0.482 & \textbf{0.661} & 0.241 & 0.753 & 0.362 \\

\textit{Flow+DE} & 256 & 0.855 & 0.270 & 0.745 & 0.343 & 0.606 & 0.218 & 0.770 & 0.338 \\
              & 512 & 0.853 & 0.284 & 0.769 & 0.358 & 0.574 & 0.205 & 0.747 & 0.321 \\

\textit{Flow+DE+FS (0.01)} & 256 & 0.901 & 0.363 & 0.886 & 0.452 & 0.626 & 0.227 & 0.774 & 0.349 \\
                           & 512 & 0.902 & 0.365 & 0.819 & 0.443 & 0.615 & 0.227 & 0.793 & 0.352 \\
\textit{Flow+DE+FS (0.02)} & 256 & 0.917 & 0.406 & 0.878 & 0.489 & 0.637 & 0.238 & 0.808 & 0.365 \\
                           & 512 & 0.912 & 0.407 & 0.884 & 0.490 & 0.641 & 0.229 & 0.782 & 0.354 \\ \hline

\textit{Flow+Packet} & 256 & 0.771 & 0.229 & 0.961 & 0.326 & 0.633 & 0.240 & 0.811 & 0.368 \\
              & \textbf{512} & 0.768 & 0.231 & 0.853 & 0.314 & 0.561 & 0.219 & \textbf{0.880} & 0.350 \\

\textit{Flow+Packet+FS (0.01)} & 256 & 0.759 & 0.234 & 0.974 & 0.331 & 0.635 & 0.244 & 0.804 & 0.372 \\
                               & \textbf{512} & 0.775 & 0.264 & \textbf{0.976} & 0.362 & 0.636 & 0.253 & 0.819 & 0.385 \\
\textit{Flow+Packet+FS (0.02)} & 256 & 0.832 & 0.281 & 0.966 & 0.380 & 0.653 & 0.240 & 0.760 & 0.364 \\
                               & 512 & 0.839 & 0.302 & 0.952 & 0.399 & 0.651 & 0.251 & 0.777 & 0.377 \\

\textit{Flow+Packet+DE} & 256 & 0.840 & 0.289 & 0.914 & 0.383 & 0.629 & 0.242 & 0.796 & 0.370 \\
                 & 512 & 0.711 & 0.174 & 0.786 & 0.247 & 0.589 & 0.224 & 0.810 & 0.350 \\

\textit{Flow+Packet+DE+FS (0.01)} & 256 & 0.744 & 0.225 & 0.946 & 0.318 & 0.604 & 0.229 & 0.837 & 0.359 \\
                                  & \textbf{512} & 0.724 & 0.207 & 0.920 & 0.299 & 0.641 & \textbf{0.255} & 0.818 & \textbf{0.388} \\
\textit{Flow+Packet+DE+FS (0.02)} & 256 & 0.854 & 0.338 & 0.889 & 0.421 & 0.615 & 0.237 & 0.868 & 0.371 \\
                                  & 512 & 0.858 & 0.330 & 0.949 & 0.427 & 0.625 & 0.231 & 0.798 & 0.358 \\
\bottomrule
\end{tabular}%
}
\vspace{0.1em} 
\parbox{\linewidth}{\scriptsize
\textit{Abbreviations:} DE: Derived Features, FS: Feature Importance Selection
}
\vspace{-14pt}
\end{table}


\subsection {
RQ3. 
\siam vs. SOTA methods for Unknown Attack Detection under a within-dataset evaluation setting
}


\rev{We compare \siam with the five unknown-attack baselines under the unknown-attack setting of both datasets, at the small-scale and large-scale budgets defined in Section~\ref{sec:datasets}. Every method is trained and evaluated on the feature representation required by its own approach, drawn from the same sampled flows: IDS-Agent, RFG-HELAD, CLOSR, and ACGAN operate on flow-level features, ACGAN additionally grouping the packets of each sampled flow into fixed-size image representations using \emph{heiFIP}\footnote{\footnotesize\url{https://github.com/stefanDeveloper/heiFIP}}
following prior work~\cite{li2024ids}; SAFE-NID operates on the non-zero-payload packets of the sampled flows\footnote{
\rev{SAFE-NID provides two out-of-distribution safeguards over the same classifier: Mahalanobis and normalizing flow. We use normalizing flow at small scale, where insufficient packets from short-flow classes make Mahalanobis covariance non-estimable (non-finite scores), and the better-performing Mahalanobis at large scale. Normalizing flow is the original safeguard and works at both scales.}
}; and \siam uses the dataset-specific RQ2 configurations, namely flow, packet, and four derived features for \ciciot, and packet features for \iot; also, it uses the best configuration from Table~\ref{tab:rq2_avg_iot23_ciciot2023}. RQ3 uses the same splits and held-out categories as RQ2.}

\rev{\subhead{Unknown-attack detection}}
\rev{Table~\ref{tab:rq3_avg_results} reports the comparison.} Overall, the results show that \siam consistently outperforms all baseline methods in a within-dataset, unknown-attack detection setting, across both datasets, particularly in terms of U-F1. On the \ciciot dataset, \rev{in small-scale setting,} \siam achieves the highest average U-F1 \rev{(0.388)}, outperforming the strongest baseline IDS-Agent (\rev{GPT-6-luna, 0.210}) by approximately \rev{84.8}\%, and surpassing \rev{ACGAN (\num{0.182}), RFG-HELAD (\num{0.140}), CLOSR (\num{0.115}), and SAFE-NID (\num{0.081}), by \num{113.2}\%, \num{177.1}\%, \num{237.4}\%, and \num{379.0}\%, respectively. At the large-scale, \siam attains a U-F1 of \num{0.379}, exceeding the strongest large-scale baselines, CLOSR and IDS-Agent (GPT-6-luna) (both \num{0.238}), by approximately \num{59.2}\%.}

\rev{Similarly,} on the \iot dataset, 
\rev{\siam outperforms all five baselines, exceeding CLOSR (U-F1 = 0.390), SAFE-NID (U-F1 = 0.372), RFG-HELAD (U-F1 = 0.277), ACGAN (U-F1 = 0.156), and IDS-Agent (GPT-6-luna, U-F1 = 0.072) by approximately 37.4\%, 44.1\%, 93.5\%, 243.6\%, and 644.4\%, respectively. At the large-scale, SAFE-NID attains \num{0.550}, above both small-scale \siam (\num{0.536}) and large-scale \siam (\num{0.450}).}
In terms of W-F1, which reflects overall classification performance on both known and unknown attack detection, \siam also demonstrates strong competitiveness. \rev{Under small-scale setting}, \siam achieves an average W-F1 of 0.641 \rev{on \ciciot and 0.930 on \iot}, \rev{outperforming the strongest baseline CLOSR by 108.1\% and 5.1\% respectively,} 
indicating that improvements in unknown detection do not come at the expense of overall accuracy. 
\rev{At the large-scale} on \iot, although SAFE-NID achieves a higher W-F1 (0.939), \siam still attains a competitive W-F1 of \rev{0.906}. \rev{On \ciciot, large-scale \siam attains the highest W-F1 of all methods (\num{0.715}), \num{112.8}\% above the strongest large-scale baseline, CLOSR (\num{0.336}).}

\rev{\subhead{Statistical significance}}
\rev{\label{significance_test}We further test whether these differences are statistically significant using two-sided exact Wilcoxon signed-rank tests~\cite{wilcoxon1945individual, demvsar2006statistical}, which are non-parametric and therefore suited to bounded measures that may be skewed or contain zeros. The unit of analysis is the held-out unknown category: U-F1 is first averaged over the ten runs within each category, and the two methods are then paired over the same nine categories, with zero-difference pairs omitted by the test. We report $\Delta$ U-F1, the difference between the two averages in Table~\ref{tab:rq3_avg_results} (\siam minus baseline), while the test operates on the nine paired category-level differences, and adjust the five \siam-versus-baseline $p$ values of each setting using Holm's method. We use the exact test because the number of pairs is small. At 100 labeled samples per known class on \ciciot, \siam is significantly better than all five baselines, by 0.307 U-F1 over SAFE-NID, 0.273 over CLOSR, 0.248 over RFG-HELAD, 0.206 over ACGAN, and 0.178 over IDS-Agent (GPT-6-luna) (Holm-adjusted $p = 0.020$ in each case). On \iot, \siam is significantly better than IDS-Agent (GPT-6-luna; $\Delta = 0.464$), ACGAN ($\Delta = 0.380$), RFG-HELAD ($\Delta = 0.259$) and CLOSR ($\Delta = 0.146$), again at $p = 0.020$, but its advantage over SAFE-NID is not significant ($\Delta = 0.164$, $p = 0.164$): the average gap is larger than the one over CLOSR, yet it is not consistent, with SAFE-NID ahead on three of the nine categories, whereas \siam leads CLOSR on all nine. The advantage of \siam is therefore statistically supported against all five baselines on \ciciot, and against four of the five on \iot.}

\rev{\subhead{Effect of training scale}}
\rev{Comparing the small-scale and large-scale settings shows that the impact of additional training data varies across methods. On \iot, SAFE-NID's U-F1 increases by approximately 47.8\% (0.372 to 0.550), and ACGAN's by 128.8\% (0.156 to 0.357). In contrast, RFG-HELAD's U-F1 decreases by 26.0\% (0.277 to 0.205), with unknown recall reaching 0.991 at a precision of only 0.126. CLOSR's U-F1 decreases by \num{4.4}\% (\num{0.390} to \num{0.373}), and its W-F1 decreases from \num{0.885} to \num{0.839}. IDS-Agent (GPT-6-luna) remains the weakest baseline on \iot\ at either budget, although its U-F1 increases by \num{29.2}\% (\num{0.072} to \num{0.093}).
For \siam, large-scale training lowers U-F1 by approximately 16.0\% (0.536 to 0.450) and W-F1 from 0.930 to 0.906, making the small-scale setting preferable on \iot.}

\rev{On \ciciot, moving from small- to large-scale increases SAFE-NID's U-F1 by \num{134.6}\% (\num{0.081} to \num{0.190}), RFG-HELAD's by \num{40.0}\% (\num{0.140} to \num{0.196}), IDS-Agent by \num{13.3}\% (GPT-6-luna, \num{0.210} to \num{0.238}), CLOSR's by \num{107.0}\% (\num{0.115} to \num{0.238}) and ACGAN only marginally from \num{0.182} to \num{0.186}. 
For \siam, large-scale training changes U-F1 little (\num{0.388} to \num{0.379}, approximately \num{2.3}\% lower) while raising W-F1 from \num{0.641} to \num{0.715}. Therefore, additional training data benefits the methods unevenly, while \siam\ trained in the small-scale setting remains the highest-scoring method on \ciciot, and \siam\ trained in the large-scale setting ranks second, ahead of every baseline at either budget}.

\rev{\subhead{Comparing the two datasets}}
Unknown-attack detection is more challenging on \ciciot: \rev{at the small-scale budget four of the six methods score a lower U-F1} than on \iot. \rev{The exceptions are ACGAN (\num{0.182} against \num{0.156}) and IDS-Agent (GPT-6-luna, \num{0.210} against \num{0.072}), which are weak on either dataset.}
This difference \rev{reflects the size of the known-class space:} 
\rev{\ciciot\ requires separating \num{25} known categories drawn from a finer-grained attack taxonomy, against five in \iot. At this budget, \siam rejects \num{26.8}\% of known \ciciot traffic as unknown, against \num{5.9}\% on \iot.}

\rev{\subhead{Category-level behavior}}
We further observe clear dataset-dependent behaviors at the category level\rev{, reported per category in 
Table~\ref{tab:per_category_uf1}}.
On \ciciot, \rev{the U-F1 of every baseline stays below \num{0.27} on every unknown category (\num{0.024} to \num{0.265}), and \siam\ is the best method on all nine, with a narrow range of \num{0.337} to \num{0.419}},
suggesting that unknown attacks are consistently hard to distinguish across categories. On \iot, several baselines struggle on specific categories (e.g., C\&C-HB-Att, C\&C-Mirai, and Okiru-Att), where U-F1 drops sharply\rev{, to \num{0.000} for SAFE-NID and IDS-Agent on C\&C-HB-Att and C\&C-Mirai despite dataset averages of \num{0.372} and \num{0.072}}. 
Despite these challenges, \siam maintains substantially higher U-F1 scores on these difficult categories 
(e.g., 0.221 on Okiru-Att and 0.264 on C\&C-Mirai), 
indicating \rev{effectiveness} in detecting unknown attacks.

\rev{\subhead{Effect of input modality}}
\rev{Finally, we examine whether the richer input representation alone explains these gains. We compare the single-modality configurations of Table~\ref{tab:rq2_avg_iot23_ciciot2023}, which use only the modality a given baseline consumes (flow- or packet-level features), against those baselines \textit{at the same small-scale budget}. On \ciciot, the best flow-only \siam configuration attains a U-F1 of \num{0.362} against \num{0.210} for the strongest flow-input baseline, IDS-Agent (GPT-6-luna), and the best packet-only configuration attains \num{0.310} against \num{0.081} for packet-input SAFE-NID, improvements of approximately \num{72.4}\% and \num{282.7}\% respectively. On \iot, the flow-only configuration attains \num{0.482} against \num{0.390} for the strongest flow-input baseline, CLOSR, and the packet-only configuration attains \num{0.536} against \num{0.372} for SAFE-NID, improvements of approximately \num{23.6}\% and \num{44.1}\%. Multimodal input is therefore not the sole explanation for the observed differences.}

\begin{summarybox}{Answer to RQ3}
The results show that \siam consistently outperforms SOTA baselines for unknown attack detection, on both datasets, in a within-dataset setting, achieving higher unknown F1-scores. 
\rev{
At small-scale, \siam has the highest U-F1 on both datasets. At large-scale, it leads on \ciciot, while SAFE-NID leads on \iot. The small-scale category-paired tests support \siam's advantage over all five baselines on \ciciot and over four of five on \iot; its difference from SAFE-NID on \iot is not significant. At large scale, \siam attains the highest U-F1 and W-F1 on \ciciot.}

\end{summarybox}

\begin{table}[t]
\centering
\caption{\rev{
Within-dataset unknown-attack detection under two training budgets (RQ3). 
}}
\label{tab:rq3_avg_results}
\vspace{-0.1in}
\setlength{\tabcolsep}{2pt}
\resizebox{\columnwidth}{!}{%
\rev{\begin{tabular}{l l cccc cccc}
\toprule
\multirow{2}{*}{\textbf{Model}} & \multirow{2}{*}{\textbf{Budget}} & \multicolumn{4}{c}{\textbf{\iot}} & \multicolumn{4}{c}{\textbf{\ciciot}} \\
\cmidrule(lr){3-6}\cmidrule(lr){7-10}
 & & \textbf{U-Pr} & \textbf{U-Rc} & \textbf{U-F1} & \textbf{W-F1} & \textbf{U-Pr} & \textbf{U-Rc} & \textbf{U-F1} & \textbf{W-F1} \\
\midrule
\multirow{2}{*}{\textit{ACGAN}} & 100 & 0.091 & \textbf{1.000} & 0.156 & 0.027 & 0.102 & 0.991 & 0.182 & 0.018 \\
 & 10\% & 0.441 & 0.539 & 0.357 & 0.561 & 0.106 & 0.764 & 0.186 & 0.065 \\
\cmidrule{1-10}
\multirow{2}{*}{\textit{CLOSR}} & 100 & 0.308 & 0.837 & 0.390 & 0.885 & 0.137 & 0.103 & 0.115 & 0.308 \\
 & 10\% & 0.265 & \textbf{1.000} & 0.373 & 0.839 & 0.136 & 0.940 & 0.238 & 0.336 \\
\cmidrule{1-10}
\multirow{2}{*}{\textit{SAFE-NID}} & 100 & 0.380 & 0.428 & 0.372 & 0.367 & 0.113 & 0.064 & 0.081 & 0.262 \\
 & 10\% & 0.471 & 0.834 & \textbf{0.550} & \textbf{0.939} & 0.146 & 0.273 & 0.190 & 0.304 \\
\cmidrule{1-10}
\multirow{2}{*}{\textit{IDS-Agent}} & 100 & 0.173 & 0.061 & 0.072 & 0.832 & 0.124 & 0.691 & 0.210 & 0.268 \\
 & 10\% & 0.081 & 0.169 & 0.093 & 0.853 & 0.163 & 0.442 & 0.238 & 0.313 \\
\cmidrule{1-10}
\multirow{2}{*}{\textit{RFG-HELAD}} & 100 & 0.222 & 0.792 & 0.277 & 0.196 & 0.104 & 0.424 & 0.140 & 0.023 \\
 & 10\% & 0.126 & 0.991 & 0.205 & 0.179 & 0.108 & \textbf{0.998} & 0.196 & 0.025 \\
\cmidrule{1-10}
\multirow{2}{*}{\textit{\siam}} & 100 & \textbf{0.484} & 0.803 & 0.536 & 0.930 & \textbf{0.255} & 0.818 & \textbf{0.388} & 0.641 \\
 & 10\% & 0.451 & 0.656 & 0.450 & 0.906 & 0.252 & 0.760 & 0.379 & \textbf{0.715} \\
\bottomrule
\end{tabular}}}

\vspace{0.3em}
\vspace{-15pt}
\end{table}

\begin{table}[t]
\centering
\caption{\rev{Cross-dataset unknown attack detection under two training budgets (RQ4). 
}}
\label{tab:cross_dataset_avg_results}
\vspace{-0.1in}
\setlength{\tabcolsep}{2pt}
\resizebox{\columnwidth}{!}{%
\rev{\begin{tabular}{l l cccc cccc}
\toprule
\multirow{2}{*}{\textbf{Model}} & \multirow{2}{*}{\textbf{Budget}} & \multicolumn{4}{c}{\textbf{\iot $\rightarrow$ \ciciot}} & \multicolumn{4}{c}{\textbf{\ciciot $\rightarrow$ \iot}} \\
\cmidrule(lr){3-6}\cmidrule(lr){7-10}
 & & \textbf{U-Pr} & \textbf{U-Rc} & \textbf{U-F1} & \textbf{W-F1} & \textbf{U-Pr} & \textbf{U-Rc} & \textbf{U-F1} & \textbf{W-F1} \\
\midrule
\multirow{2}{*}{\textit{ACGAN}} & 100 & 0.214 & 0.989 & 0.351 & 0.078 & 0.043 & \textbf{1.000} & 0.079 & 0.007 \\
 & 10\% & 0.689 & 0.344 & 0.421 & 0.518 & 0.299 & 0.565 & 0.229 & 0.081 \\
\cmidrule{1-10}
\multirow{2}{*}{\textit{CLOSR}} & 100 & 0.436 & 0.723 & 0.535 & 0.738 & 0.271 & 0.815 & 0.351 & 0.349 \\
 & 10\% & 0.510 & 0.960 & 0.666 & 0.792 & 0.024 & 0.372 & 0.043 & 0.334 \\
\cmidrule{1-10}
\multirow{2}{*}{\textit{SAFE-NID}} & 100 & \textbf{0.864} & 0.568 & 0.675 & 0.414 & 0.158 & 0.276 & 0.191 & 0.219 \\
 & 10\% & 0.751 & 0.611 & 0.672 & 0.857 & 0.156 & 0.709 & 0.232 & 0.317 \\
\cmidrule{1-10}
\multirow{2}{*}{\textit{IDS-Agent}} & 100 & 0.685 & 0.100 & 0.161 & 0.712 & 0.047 & 0.612 & 0.083 & 0.272 \\
 & 10\% & 0.440 & 0.111 & 0.177 & 0.742 & 0.054 & 0.496 & 0.085 & 0.314 \\
\cmidrule{1-10}
\multirow{2}{*}{\textit{RFG-HELAD}} & 100 & 0.577 & 0.638 & 0.573 & 0.221 & \textbf{0.341} & 0.745 & \textbf{0.383} & 0.029 \\
 & 10\% & 0.297 & \textbf{0.991} & 0.457 & 0.195 & 0.047 & \textbf{1.000} & 0.086 & 0.007 \\
\cmidrule{1-10}
\multirow{2}{*}{\textit{\siam}} & 100 & 0.802 & 0.863 & \textbf{0.816} & \textbf{0.913} & 0.120 & 0.857 & 0.192 & 0.657 \\
 & 10\% & 0.605 & 0.258 & 0.360 & 0.782 & 0.123 & 0.863 & 0.196 & \textbf{0.738} \\
\bottomrule
\end{tabular}}}

\vspace{0.3em}
\vspace{-15pt}
\end{table}

\begin{table*}[!htbp]
\centering
\caption{\rev{U-F1 per held-out unknown category, small-scale within-dataset setting (RQ3). Means over the same ten runs.}}

\label{tab:per_category_uf1}
\vspace{-0.1in}
\setlength{\tabcolsep}{2.5pt}
\resizebox{\textwidth}{!}{%
\rev{\begin{tabular}{l *{9}{c} c *{9}{c} c}
\toprule
\multirow{2}{*}{\textbf{Model}} & \multicolumn{10}{c}{\textbf{\ciciot}} & \multicolumn{10}{c}{\textbf{\iot}} \\
\cmidrule(lr){2-11}\cmidrule(lr){12-21}
 & \textbf{Backdoor} & \textbf{Hijack} & \textbf{CmdInj} & \textbf{DDoS} & \textbf{DictBrute} & \textbf{Recon} & \textbf{SqlInj} & \textbf{Upload} & \textbf{XSS} & \textbf{Avg}
 & \textbf{C\&C-FD} & \textbf{C\&C-HB} & \textbf{C\&C-HB-Att} & \textbf{C\&C-HB-FD} & \textbf{C\&C-Mirai} & \textbf{C\&C-Torii} & \textbf{FileDL} & \textbf{Okiru} & \textbf{Okiru-Att} & \textbf{Avg} \\
\midrule
\textit{ACGAN} & 0.182 & 0.183 & 0.182 & 0.182 & 0.182 & 0.182 & 0.182 & 0.182 & 0.182 & 0.182 & 0.346 & 0.346 & 0.010 & 0.104 & 0.021 & 0.060 & 0.160 & 0.346 & 0.011 & 0.156 \\
\textit{CLOSR} & 0.056 & 0.265 & 0.054 & 0.132 & 0.244 & 0.101 & 0.071 & 0.032 & 0.082 & 0.115 & 0.523 & 0.819 & 0.091 & 0.349 & 0.167 & 0.362 & 0.456 & 0.672 & 0.072 & 0.390 \\
\textit{SAFE-NID} & 0.143 & 0.037 & 0.052 & 0.082 & 0.114 & 0.024 & 0.106 & 0.070 & 0.103 & 0.081 & \textbf{0.934} & 0.000 & 0.000 & \textbf{0.849} & 0.000 & 0.185 & \textbf{0.835} & 0.339 & 0.211 & 0.372 \\
\textit{IDS-Agent} & 0.204 & 0.207 & 0.165 & 0.228 & 0.236 & 0.222 & 0.211 & 0.205 & 0.208 & 0.210 & 0.117 & 0.058 & 0.000 & 0.137 & 0.000 & 0.089 & 0.161 & 0.034 & 0.050 & 0.072 \\
\textit{RFG-HELAD} & 0.171 & 0.172 & 0.140 & 0.061 & 0.165 & 0.132 & 0.144 & 0.135 & 0.143 & 0.140 & 0.644 & 0.437 & 0.049 & 0.321 & 0.061 & 0.214 & 0.423 & 0.319 & 0.025 & 0.277 \\
\textit{\siam} & \textbf{0.405} & \textbf{0.419} & \textbf{0.410} & \textbf{0.397} & \textbf{0.376} & \textbf{0.337} & \textbf{0.368} & \textbf{0.387} & \textbf{0.393} & 0.388 & 0.811 & \textbf{0.832} & \textbf{0.163} & 0.649 & \textbf{0.264} & \textbf{0.502} & 0.695 & \textbf{0.690} & \textbf{0.221} & 0.536 \\
\bottomrule
\end{tabular}}}
\vspace{-15pt}
\end{table*}



\vspace{-0.1in}
\subsection{RQ4. 
\siam vs. SOTA methods for Unknown Attack Detection under a cross-dataset evaluation setting
}


\rev{We reuse the trained models in RQ3 
and evaluate them in a cross-dataset setting (train$\rightarrow$test), in which known and benign test samples are drawn from the \textit{train} dataset while unknown-attack samples come from the \textit{test} dataset. Models trained on \ciciot are therefore evaluated on \num{450} known/benign samples from \ciciot together with \num{189} unknown samples from \iot; models trained on \iot are evaluated on \num{189} known/benign samples from \iot together with \num{450} unknown samples from \ciciot. Table~\ref{tab:cross_dataset_avg_results} reports the results under the small-scale and large-scale settings.}
Statistical significance is assessed as in RQ3, with the five \siam-versus-baseline comparisons of each transfer direction forming one Holm family.

\rev{\subhead{Small-scale setting}}
\textit{(i) \iot $\rightarrow$ \ciciot:} 
\siam achieves the best overall performance among all methods, with a U-F1 of 0.816 and a W-F1 of 0.913. This substantially outperforms the strongest baseline SAFE-NID (U-F1 = \rev{0.675}) by approximately \rev{20.9}\%, and exceeds \rev{RFG-HELAD (\num{0.573}), CLOSR (\num{0.535}), ACGAN (\num{0.351}), and IDS-Agent (GPT-6-luna, \num{0.161}) by \num{42.4}\%, \num{52.5}\%, \num{132.5}\%, and \num{406.8}\%, respectively. The reported category-paired tests mark all five U-F1 differences as significant (Holm-adjusted \(p=\num{0.020}\) in each case)}. 
\rev{
In terms of W-F1, \siam attains 0.913 against a highest small-scale baseline value of 0.738 for CLOSR, an improvement of approximately 23.7\%; SAFE-NID, the next baseline, reaches 0.414.} Notably, although RFG-HELAD attains an extremely high unknown recall (U-Rc = 0.991), its poor precision leads to a significantly lower U-F1, whereas \siam maintains a more balanced precision-recall trade-off. These results indicate that \siam generalizes effectively across datasets when trained on one dataset (e.g., \iot) and tested on a more complex dataset (e.g., \ciciot).

\textit{\noindent(ii) \ciciot $\rightarrow$ \iot:}
The overall performance of all methods drops markedly. While \siam still achieves the highest W-F1 (0.657), its U-F1 score decreases to 0.192, which is comparable to or lower than that of several baseline methods, including SAFE-NID (\rev{0.191; $\Delta = +0.001$, $p = 0.820$}). \rev{CLOSR and RFG-HELAD obtain higher U-F1 scores of 0.351 and 0.383, respectively. When performance is compared across the nine held-out unknown categories, the difference between \siam and CLOSR is statistically significant ($\Delta=-0.159$, $p = 0.035$), favouring CLOSR. Although RFG-HELAD also attains a higher U-F1, its difference from \siam is not statistically significant ($\Delta=-0.191$, $p = 0.148$). Against ACGAN ($\Delta = +0.113$) and IDS-Agent (GPT-6-luna; $\Delta = +0.109$), \siam remains significantly better in this direction ($p = 0.020$ in both cases).
}
Overall, these results suggest that unknown attack detection becomes more challenging when models are trained on \ciciot and evaluated on \iot, even for \siam. Nevertheless, \siam consistently maintains higher unknown recall (U-Rc = 0.857) than most baselines, indicating it remains sensitive to unseen attacks, albeit at the cost of reduced precision in this setting.

\rev{\subhead{Large-scale setting} 
Increasing the training sample size leads to substantial but direction-dependent changes in unknown-attack detection performance, with the effect depending on both the method and the transfer direction. 
\textit{(i) \iot $\rightarrow$ \ciciot:} 
U-F1 is nearly unchanged for SAFE-NID (0.675 to 0.672) and improves for CLOSR by approximately 24.5\% (0.535 to 0.666), the largest gain of any method in this direction, for ACGAN by 19.9\% (0.351 to 0.421) and for IDS-Agent by 9.9\% (0.161 to 0.177). RFG-HELAD decreases by approximately 20.2\% (0.573 to 0.457). For \siam, U-F1 decreases by approximately \num{55.9}\% (\num{0.816} to \num{0.360}), caused by a drop in unknown recall from \num{0.863} to \num{0.258}.
\textit{\noindent(ii) \ciciot $\rightarrow$ \iot:} U-F1 improves for ACGAN by approximately 189.9\% (0.079 to 0.229) and for SAFE-NID by 21.5\% (0.191 to 0.232), and changes little for IDS-Agent (2.4\%, 0.083 to 0.085) and for \siam (\num{0.192} to \num{0.196}), whose W-F1 nevertheless rises from \num{0.657} to \num{0.738}, the highest of all methods. RFG-HELAD decreases by approximately 77.5\% (0.383 to 0.086), with an unknown recall of 1.000 at an unknown precision of only 0.047, so its high recall comes at the cost of a very high false-positive rate, the fraction of known and benign samples rejected as unknown. CLOSR falls furthest, by approximately 87.7\% (0.351 to 0.043), and here the decline is not a precision--recall trade-off: unknown recall falls to 0.372 alongside an unknown precision of 0.024, so neither quantity is retained. IDS-Agent changes least overall, by at most 9.9\% in either direction.}

\rev{These results provide a common explanation. The additional training data is taken from the source dataset only. It describes the dataset's known classes more completely, but says nothing about the unknown attacks used at test time, which come from the other dataset. Its effect is therefore seen in the balance between unknown precision and unknown recall: for most methods one of the two rises while the other falls, so a larger training set does not reliably increase U-F1. Since the source dataset is different in each transfer direction, the same increase in training data affects this balance differently in each. Additional data increases U-F1 in both directions for ACGAN and IDS-Agent, lowers it in both for RFG-HELAD, and helps one direction while harming the other for CLOSR and \siam; for SAFE-NID it leaves the \iot $\rightarrow$ \ciciot result almost unchanged and improves the reverse.}

\subhead{Cross- vs within-dataset} Clear differences emerge between cross- and within-dataset performance on the same test data.

\noindent \textit{On the \ciciot unknown-attack test set \rev{at the small-scale},} \siam achieves a much higher U-F1 under cross-dataset setting (\iot $\rightarrow$ \ciciot, U-F1 = 0.816) than within-dataset setting (U-F1 = 0.388, \rev{Table~\ref{tab:rq3_avg_results}}), \rev{an increase of approximately 110.3\%,}
suggesting training on \iot yields a representation space in which \ciciot attacks are more easily separated from known traffic, improving unknown-attack detection.
\textit{On the \iot unknown-attack test set \rev{at the small-scale},} cross-dataset performance (\ciciot $\rightarrow$ \iot, U-F1 = 0.192) is substantially worse than within-dataset performance (U-F1 = 0.536,
\rev{Table~\ref{tab:rq3_avg_results}}), \rev{a decrease of approximately 64.2\%}. A similar degradation trend can be observed for most baseline methods. 

\rev{\noindent\textit{On the \ciciot unknown-attack test set at the large-scale,} the comparison reverses: cross-dataset U-F1 (\num{0.360}) falls approximately \num{5.0}\% below the within-dataset value (\num{0.379}, Table~\ref{tab:rq3_avg_results}), so the advantage of training on \iot\ disappears once its training set is enlarged. On \iot, cross-dataset performance (\num{0.196}) stays approximately \num{56.4}\% below within-dataset performance (\num{0.450}), as at the small-scale budget.}

\rev{The observed asymmetry between the two cross-dataset directions can be attributed to the differences in the source-domain representations and their resulting rejection behavior. When trained on \ciciot, \siam learns a broader known-attack representation that rejects a larger fraction of known traffic as unknown (26.8\%, compared with 5.9\% in the within-dataset setting of RQ3). Consequently, when this model is transferred to \iot, the broad rejection behavior persists (false-positive rate of 26.7\%), causing many samples to be classified as unknown and reducing unknown precision. In contrast, training on \iot produces a tighter known-attack representation with a much lower false-positive rate (6.3\% after transfer), allowing \ciciot attack samples to be rejected as unknown with considerably higher precision. This is therefore driven by how the source dataset shapes the model's known-attack representation and rejection behavior, rather than by threshold calibration. 
The larger and finer-grained known-class space of \ciciot produces
this broader representation, whereas the more compact attack distribution of \iot results in a tighter known-attack manifold.}

\begin{summarybox}{Answer to RQ4}
Overall, \siam demonstrates cross-dataset \rev{effectiveness} compared to SOTA methods, particularly when trained on \iot and tested on the \ciciot dataset. 
\rev{
At small-scale, \siam shows the highest U-F1 in the \iot\(\rightarrow\)\ciciot direction; in the reverse direction, it is nearly equal to SAFE-NID and below RFG-HELAD and CLOSR. At large scale, SAFE-NID has the highest U-F1 in both directions, while \siam retains the highest W-F1 in the \ciciot\(\rightarrow\)\iot direction. Cross-dataset performance is therefore direction- and budget-dependent.}
\end{summarybox}

\vspace{-0.1in}
\section{Discussion}
\label{sec:discussion}
\subsection{Findings and Implications}

\subhead{\rev{BERT-based models} scale better}
The results in RQ1 reveal clear differences in data efficiency and scalability among ML, DL, and \rev{BERT-based} models. 
ML models, particularly tree-based ensembles, remain competitive in small-scale settings because they rely on explicit feature engineering, whereas DL models struggle with limited data. In contrast, \rev{BERT-based} models benefit from pretrained representations and exhibit superior robustness and scalability when sufficient labeled data is available.

\subhead{Feature matters}
RQ2 shows that the performance of unknown-attack detection is driven mainly by feature composition rather than \rev{the maximum token length}. Combining packet-level and flow-level features consistently improves U-F1, \rev{with the largest gain on \ciciot, where it adds \num{4}--\num{6}\% over the best single-modality configuration; \ciciot\ requires separating \num{25} known categories, against five in \iot}.
Feature importance selection further enhances robustness by removing noise, 
\rev{whereas increasing the token length} yields only marginal and inconsistent gains.

\rev{\subhead{Pretraining matters}}
\rev{Pretraining improves unknown-attack detection on both datasets. Cybersecurity-domain pretraining provides an additional benefit over general-domain pretraining on \ciciot, but not measurably on \iot. The input representation is dataset dependent: the serialized pretrained encoder performs best on \ciciot, whereas direct numeric encoding with the Transformer performs best on \iot. Thus, the serialized pretrained encoder is most beneficial for the dataset with a larger known-class space, while a tabular model remains a strong alternative when the known-class space is smaller.}



\subhead{Dataset bias matters}
\rev{RQ4 reveals direction-dependent cross-dataset transfer. Transfer from \iot\ to \ciciot\ is effective, whereas transfer from \ciciot\ to \iot\ yields high unknown recall but low unknown precision, resulting in frequent false positives. Models trained on \ciciot\ reject approximately \num{27}\% of known traffic as unknown both within \ciciot\ and after transfer to \iot, compared with approximately \num{6}\% for models trained on \iot. The weaker \ciciot-to-\iot\ transfer therefore reflects the rejection behavior learned from the source data. 
This behavior is consistent with the known-class spaces of the two datasets: \ciciot\ contributes \num{25} known categories 
drawn from a finer-grained attack taxonomy, against five in \iot, producing a broader known-class representation that rejects more \iot\ traffic as unknown. Conversely, the more compact \iot\ attack distribution transfers more effectively to \ciciot.
}

\subsection{Threats to Validity}
\noindent(i) \textit{Internal validity:} Implementation, configuration, feature granularity, and randomness may affect results. Authors' baseline implementations were sometimes incomplete or dataset-specific; we reimplemented RFG-HELAD's hardcoded threshold selection from its documentation and validated against published results~\cite{zhong2024rfg}. IDS-Agent uses GPT-6-luna throughout; repeating \num{10}\% runs with GPT-4o changes within-dataset U-F1 by at most \num{0.021} \tnew{(GPT-4o versus GPT-6-luna: \num{0.104} versus \num{0.093} on \iot, \num{0.217} versus \num{0.238} on \ciciot)}, leaving the within-dataset conclusions unchanged\tnew{; under cross-dataset transfer GPT-6-luna is better in both directions (\num{0.052} versus \num{0.177} from \iot\ to \ciciot, \num{0.062} versus \num{0.085} in reverse), so reporting GPT-4o would understate this baseline}.
We average ten runs using identical splits and protocols across methods.
(ii) \textit{External validity:} Two public IoT datasets differ in attack taxonomy, scale, and traffic characteristics but cannot represent deployment diversity; generalization to other environments, proprietary traffic, or emerging attacks requires larger, operational datasets.
(iii) \textit{Construct validity:} Held-out known categories proxy true zero-days and may yield optimistic unknown-detection estimates because they share the known classes' capture; RQ4 mitigates this using cross-dataset unknowns.
(iv) \textit{Conclusion validity:} Small-scale means over ten runs undergo Holm-corrected paired tests across nine held-out categories, with non-significant differences identified. Single-run large-scale results provide neither variance estimates nor support for significance claims.

\section{Conclusion}\label{sec:conclusion}

In this paper, we systematically investigated IoT attack detection under both known- and unknown-attack settings. Specifically, we (i) conducted a comprehensive evaluation of ML-, DL-, and \rev{BERT-based} models for IoT attack detection, and (ii) proposed a novel Siamese network and SecBERT-based approach, \siam, for unknown attack detection under both within- and cross-dataset scenarios, on two IoT datasets. Our experimental results revealed that while traditional ML models remain competitive in data-scarce settings, \rev{BERT-based models provide stronger scalability and robustness for known IoT attack detection as more labeled data become available.}
\rev{For unknown-attack detection, \siam\ achieves the highest within-dataset unknown F1-score on both datasets using \num{100} labeled samples per known class, with improvements that are statistically significant across ten runs against all five baseline methods on one dataset and against four of the five on the other. Its cross-dataset results, however, differ by transfer direction and by label budget: the weaker \ciciot-to-\iot\ transfer is associated with the high false-positive rate learned from the \ciciot\ source data, rather than threshold calibration. The controlled encoder--representation analysis further shows that pretrained encoders improve detection on both datasets, cybersecurity-domain pretraining adds value on \ciciot, and direct numeric encoding with the Transformer remains competitive on \iot.}
In future work, we plan to evaluate our approach on a broader range of IoT datasets, including larger-scale, more diverse, and real-world traffic traces, to gain deeper insights into its robustness under realistic network conditions and evolving attack behaviors. We will also extend \siam with additional modalities, such as temporal traffic dynamics, protocol semantics, and device-level context, to enhance representation learning and further improve the performance of unknown attack detection in heterogeneous IoT environments.
\vspace{-0.1in}

\bibliographystyle{IEEEtran}
\bibliography{mybib}


\clearpage

\vfill

\end{document}